\documentclass[prb, superscriptaddress ,
,showpacs,preprintnumbers,amsmath,amssymb,floatfix]{revtex4}
\usepackage{color}
\usepackage{graphicx}
\usepackage{dcolumn}
\usepackage{bm}
\usepackage[mathscr]{eucal}
\usepackage[nooneline]{subfigure}
\usepackage{amsmath} 

\newcommand{\ud}{\mathrm{d}}

\newcommand{\bra}[1]{\langle#1\vert}
\newcommand{\ket}[1]{\vert#1\rangle}
\newcommand{\op}[2]{\ket{#1}\bra{#2}}

\newcommand{\eqn}[1]{Eq.~(\ref{#1})}
\newcommand{\eqns}[1]{Eqs.~(\ref{#1})}
\newcommand{\ignore}[1]{}

\newcommand{\commute}[2]{[#1,#2]}

\DeclareMathOperator{\Tr}{Tr} \DeclareMathOperator{\U}{H}
\newcommand{\tr}[1]{\Tr\left\{#1\right\}}
\newcommand{\fig}[1]{Fig.~\ref{#1}}
\newcommand{\figs}[1]{Figs.~\ref{#1}}

\DeclareMathOperator{\ramp}{\Theta}
\DeclareMathOperator{\prob}{Prob}

\newcommand{\hc}{\mathsf{H.c.}}

\newcommand{\Boltz}{k_B}

\newcommand{\splitting}{\phi}
\newcommand{\trel}{\tau_\mathrm{r}}
\newcommand{\tdeph}{\tau_\mathrm{d}}
\newcommand{\tmeas}{\tau_\mathrm{m}}

\newcommand{\hsys}{H_\mathrm{sys}}

\newcommand{\hL}{H_\source}
\newcommand{\hR}{H_\drain}

\newcommand{\hmeas}{H_\mathrm{meas}}
\newcommand{\hrot}{H_\mathrm{0}}
\newcommand{\hi}{H_\mathrm{I}}
\newcommand{\htot}{H_\mathrm{Tot}}

\newcommand{\hleads}{H_\mathrm{leads}}

\newcommand{\rhoi}{\rho_\mathrm{I}}
\newcommand{\rhos}{\rho}

\newcommand{\dummy}{t'}
\newcommand{\lind}{\mathcal{D}}
\newcommand{\jump}{{\mathcal J}}
\newcommand{\A}{\mathcal{A}}
\newcommand{\Tnd}{\mathcal{T}}
\newcommand{\too}{T_{00}}
\newcommand{\xoo}{\chi_{00}}
\newcommand{\xgm}{\mathcal{X}}
\newcommand{\Lind}{\mathcal{L}}
\newcommand{\ghb}{G_{\mathrm{hb}}}
\newcommand{\shb}{S_{\mathrm{hb}}}
\newcommand{\glb}{G_{\mathrm{lb}}}
\newcommand{\slb}{S_{\mathrm{lb}}}
\newcommand{\ggm}{G_{\mathrm{GM}}}
\newcommand{\sgm}{S_{\mathrm{GM}}}
\newcommand{\LT}{\textsf{L}}
\newcommand{\FT}{\textsf{F}}

\newcommand{\source}{S}
\newcommand{\drain}{D}
\newcommand{\current}{I}
\newcommand{\iloc}{\bar \current_\mathrm{loc}}
\newcommand{\diloc}{\delta \current_\mathrm{loc}}
\newcommand{\tms}[1]{{#1}}
\newcommand{\del}[1]{}

\begin{document}

\title{Continuous measurement of a charge qubit with a point
contact detector at arbitrary bias: the role of inelastic
tunnelling}

\author{T. M. Stace}
\affiliation{Cavendish Laboratory, University of Cambridge,
Madingley Road, Cambridge CB3 0HE, UK} \email{tms29@cam.ac.uk}
\author{S. D. Barrett}
\affiliation{Cavendish Laboratory, University of Cambridge,
Madingley Road, Cambridge CB3 0HE, UK} \affiliation{Hewlett
Packard Laboratories,  Filton Road, Stoke Gifford Bristol, BS34
8QZ} \email{sean.barrett@hp.com}
\date{\today}



\begin{abstract}
We study the dynamics of a charge qubit, consisting of a single
electron in a double well potential, coupled to a point-contact
(PC) electrometer using the quantum trajectories formalism. In
contrast with previous work, our analysis is valid for
\emph{arbitrary} source-drain bias across the PC, but is
restricted to the sub-Zeno \tms{limit}. We find that the dynamics
is strongly affected by inelastic tunnelling processes in the PC.
These processes reduce the efficiency of the PC as a qubit readout
device, and induce relaxation even when the source-drain bias is
zero. We show that the sub-Zeno dynamics are divided into two
regimes: low- and high-bias in which the PC current and current
power spectra show markedly different behaviour. To further
illustrate the division between the regimes and the inefficiency
of the detector, we present simulated quantum trajectories of the
conditional qubit and detector dynamics.  We also describe how
single shot measurements in an arbitrary basis may be achieved in
the sub-Zeno regime.
\end{abstract}
\maketitle

\section{Introduction}

Single shot quantum measurement of mesoscopic systems is
recognized as an important goal.  Fundamentally, it will allow us
to make time-resolved observations of  quantum mechanical effects
in such systems, and practically it will be a necessary component
in the construction of solid-state quantum information processors
(QIP). There are numerous proposals for implementing QIP in solid
state systems, using doped silicon \cite{kan98}, electrostatically
defined quantum dots \cite{los98} and superconducting boxes
\cite{mak01}.  In these proposals, the output of the QIP is
determined by measuring the position of a single electron or
Cooper pair.  Single electrons hopping onto single quantum dots
have been observed on a microsecond time scale using
single-electron transistors (SET) \cite{lu03}. Ensemble
measurements of a double well system (qubit) have been
demonstrated in superconducting devices \cite{pas03}.  So far,
single shot qubit measurements remain elusive.

\begin{figure}
\includegraphics[width=4cm]{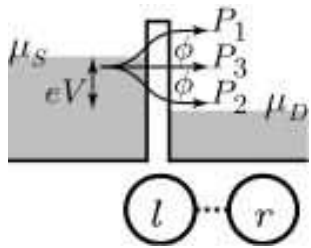}
\caption{\label{fig:schematic} Schematic of the qubit and PC
showing lead energy bands.  Electrons tunnelling from the source
to the drain may do so elastically or inelastically, depicted by
arrows.  Different transitions induce different jumps, $P_i$, on
the qubit.}
\end{figure}

It is therefore important to consider the measurement of single
electron qubits by sensitive electrometers.  Two possible
electrometers have been discussed to date: SETs (see e.g.
\citet{mak01}) and point contacts (PCs)
\cite{goa01a,goa01,gur97,kor01a,kor01b,moz02,pil02,gur03,mak00}.
PCs have been shown to be sensitive charge detectors
\cite{fie93,buk98,gar03,elz03}, and are the focus of this work.
Figure \ref{fig:schematic} illustrates the physical system we
consider here, with a PC (represented by two Fermi seas separated
by a tunnel barrier) in close proximity to one of the double well
minima.  Despite the apparent simplicity of this system it
exhibits a rich range of behaviour.

There are several energy scales of relevance: the splitting of the
qubit eigenstates, $\splitting$, the bias voltage applied across
the PC, $e V=\mu_S-\mu_D$, and the measurement induced dephasing
rate, $\Gamma_d$, due to the different currents through the PC
that result for different localized qubit electron states.  These
three energies define three distinct measurement regimes
\begin{enumerate}
\item low-bias regime, where $\Gamma_d/2 \ll e V<\splitting$.
\item high-bias regime, where $\Gamma_d/2 \ll \splitting <e V$.
\item quantum Zeno limit, where $\splitting \ll \Gamma_d/2 \ll  e
V$.
\end{enumerate}

In the quantum Zeno limit, frequent weak measurements  localize
the qubit, suppressing its dynamics \cite{goa01a,kor01b}. \tms{To
date, little attention has been paid to the low bias regime
\cite{stace:136802}.  Recently, the \emph{asymmetric} power
spectrum was calculated for arbitrary $eV$\cite{shn02}.  This may
correspond to the emission and absorption power spectra of quanta
in the \tms{PC}, as discussed in \cite{agu00}.  Unconditional
calculations of the dynamics for arbitrary $eV$ have also been
discussed recently \cite{li:085315}}.

We present a detailed analysis of the sub-Zeno limit, for
arbitrary $eV$, showing the sharp difference in behaviour between
the low- and high-bias regimes. Our analysis demonstrates the
importance of inelastic tunnelling processes in the PC, in which
electrons tunnelling through the PC barrier exchange energy with
the qubit. These inelastic processes, illustrated in Fig.
\ref{fig:schematic}, incoherently excite or relax the qubit, yet
are not in general experimentally distinguishable from elastic
tunnelling processes using current measurement.

The paper begins with a derivation of an unconditional master
equation for the joint system consisting of qubit and PC,
highlighting the significance of the Fourier decomposition of the
interaction Hamiltonian. We then present analytic solutions to
this master equation, in both the high- and low-bias regimes. In
Sec.\ \ref{sec:CME}, we derive \emph{conditional} master equations,
which describe the dynamics of the qubit conditioned on particular
realizations of the PC detector output. In Sec.\
\ref{sec:SteadyState} we apply the results of the preceding
sections to determine steady state properties such as the steady
state density matrix (which demonstrates the close analogy between
the PC bias voltage and a heat bath), and the corresponding
detector output current. Following this, we derive steady state
power spectra.  We end the analysis with some sample trajectories,
found by solving the conditional master equations, showing
possible measurement outcomes. We conclude the paper with a
discussion of the implications of our results for qubit
spectroscopy, and for quantum information processing.

\section{System Hamiltonian}

We model the double well system as a two level system, on the
basis that the two lowest energy eigenstates, $\ket{e}$ and
$\ket{g}$ of the double well system are well separated from higher
lying single particle energy eigenstates.  We write these
eigenstates as
\begin{eqnarray}
\ket{e}&=&\cos(\theta/2)\ket{r}-\sin(\theta/2)\ket{l},\nonumber\\
\ket{g}&=&\sin(\theta/2)\ket{r}+\cos(\theta/2)\ket{l},\nonumber
\end{eqnarray}
where $\ket{l}$ and $\ket{r}$ are the left and right localized
states respectively. The Hamiltonian for the double well system,
its interaction with the PC leads and the lead Hamiltonian  is
\begin{eqnarray}
\hsys&=&-\frac{\Delta}{2}\sigma_x-\frac{\epsilon}{2}\sigma_z=-\splitting\,\sigma_z^{(e)}/2,\\
\hmeas&=&\sum_{k,q} (T_{k,q}+\chi_{k,q} \sigma_z) a^\dagger_{\drain,q} a_{\source,k}+\hc\\
\hleads&=&\hL+\hR=\sum_k \omega_k (a^\dagger_{\source,k}
a_{\source,k}+a^\dagger_{\drain,k} a_{\drain,k})
\end{eqnarray}
where $\sigma_x=\op{l}{r}+\op{r}{l}$,
$\sigma_z=\op{l}{l}-\op{r}{r}$,
$\theta=\tan^{-1}(\frac{\Delta}{\epsilon})$,
$\splitting=\sqrt{\Delta^2+\epsilon^2}$ and
$\sigma_z^{(e)}=\op{g}{g}-\op{e}{e}$.  We adopt the convention
that $\chi_{k,q}<0$ so that the left well is nearest the PC.  When
$\theta=0$ the energy eigenstates coincide with the localized
states, since tunnelling is effectively switched off.  At
$\theta=\pi/2$, the tunnelling rate, $\Delta$, dominates the bias,
$\epsilon$, so the eigenstates are the completely delocalized
states.

\section{Unconditional Master Equation}

\label{sec:UME}

The von Neumann equation for the density matrix, $R$, of the bath
and system is
\begin{equation}
\dot R(t)=-i\commute{\htot}{R}.
\end{equation}
To derive the master equation for the reduced density operator,
$\rho$, of the double well system we transform to an interaction
picture with respect to the free Hamiltonian
$\hrot=\hsys+\hleads$, $\hi(t)=e^{i \hrot t}\htot e^{-i \hrot
t}-\hrot=e^{i \hrot t}\hmeas e^{-i \hrot t}$, expand the von
Neumann equation to second order and trace over the lead modes,
i.e.
\begin{equation}
\dot
\rhoi(t)=\Tr_{\source,\drain}\{-i\commute{\hi(t)}{R(0)}-\int_{0}^t
d
\dummy\commute{\hi(t)}{\commute{\hi(\dummy)}{R(\dummy)}}\},\label{eqn:SecondOrderExpansion}
\end{equation}
where
\begin{eqnarray}
\hi(t)&=&\sum_{k,q} e^{-i(\omega_k-\omega_q)t}(T_{k,q}+\chi_{k,q} \sigma_z(t))a^\dagger_{\drain,q} a_{\source,k}+\hc\equiv\sum_{k,q} S_{k,q}(t)a^\dagger_{\drain,q} a_{\source,k}+\hc,\\
\textrm{and }\sigma_z(t)&=&\begin{pmatrix}
     \cos(\theta) & e^{-i t \splitting} \sin(\theta)  \\
     e^{i t \splitting} \sin(\theta) & - \cos(\theta)
\end{pmatrix}
\end{eqnarray}
and $S_{k,q}(t)$ is a time dependent system operator that may be
written as a discrete Fourier decomposition $S_{k,q}(t)=\sum_n
e^{-i(\omega_k-\omega_q+\omega_n)t}P_n$ for some time independent
operators $P_n$ and frequencies $\omega_n$.  We can write the
operator $S_{k,q}(t)$ explicitly as
\begin{equation}
S_{k,q}(t)=e^{-i t (\omega_k-\omega_q)}(e^{-i t\splitting}P_1+e^{i
t \splitting}P_2+ P_3),\label{eqn:S}
\end{equation}
where $P_1=P_2^\dagger={\xoo \sin(\theta)}{}\op{g}{e}$, $P_3=(\too
+{\xoo \cos(\theta)}{}\sigma_z^{(e)})$  and we have assumed that
the tunnelling matrix elements $T_{k,q}=\too$ and
$\chi_{k,q}=\xoo$ are constant over the regime of interest.

The form of $S_{k,q}(t)$ indicates that there are three possible
jump processes, indicated in \fig{fig:schematic}. $P_3$ is
associated with  elastic tunnelling of electrons through the PC.
$P_1$ and $P_2$ are associated respectively with inelastic
excitation and relaxation of electrons tunnelling through the PC
accompanied by an energy transfer $\phi$.  This energy is provided
by the qubit which relaxes or excites in response.  Inelastic
transitions in similar systems have been described in
\cite{agu00}, which calculated the current power spectrum through
an \emph{open} double well system due to shot noise through a
nearby PC.

Assuming the leads are always near thermal equilibrium,
$\Tr_{\source,\drain}\{a^\dagger_{i,k}R(t)\}=\Tr_{\source,\drain}\{a_{i,k}R(t)\}=0$
and $\Tr_{\source,\drain}\{a^\dagger_{i,k}
a_{n,k}R(t)\}=\delta_{i,j}f_i(\omega_k)\rhos(t)$, where
$i,j\in\{L,R\}$ and $f_i$ is the Fermi distribution for lead $i$.
We also make the common assumption \cite{gar00} that if the lead
correlation time is much shorter than other time scales, then the
lower limit on the time integral in \eqn{eqn:SecondOrderExpansion}
may be set to $-\infty$.  Equation
(\ref{eqn:SecondOrderExpansion}) becomes
\begin{eqnarray}
\dot \rhoi(t)\approx{}&-&\int_{-\infty}^t d \dummy \int \ud\omega_k\int \ud\omega_q g_\source(\omega_k) g_\drain(\omega_q) f_\source(\omega_k)(1-f_\drain(\omega_q))\nonumber\\
&&\times\left(S(t)^\dagger S(\dummy)\rhoi(\dummy)-S(t)\rhoi(\dummy)S(\dummy)^\dagger-S(\dummy)\rhoi(\dummy)S(t)^\dagger+\rhoi(\dummy)S(\dummy)^\dagger S(t)\right)\nonumber\\
&-&\int_{-\infty}^t d \dummy \int \ud\omega_k\int \ud\omega_q g_\source(\omega_k) g_\drain(\omega_q)f_\drain(\omega_q)(1-f_\source(\omega_k))\nonumber\\
&&\times\left(S(t)
S(\dummy)^\dagger\rhoi(\dummy)-S(t)^\dagger\rhoi(\dummy)S(\dummy)-S(\dummy)^\dagger\rhoi(\dummy)S(t)+\rhoi(\dummy)S(\dummy)
S(t)^\dagger\right),\label{eqn:mastereqn1}
\end{eqnarray}
where we have made the standard replacement
$\sum_{k}\rightarrow\int \ud\omega_{k}
g_{\source(\drain)}(\omega_{k})$ where
$g_{\source(\drain)}(\omega_k)$ is the density of states for lead
$\source(\drain)$ which we will hereafter assume is a constant,
$g_{\source(\drain)}$, in the energy range of interest.

We wish to evaluate the integrals in \eqn{eqn:mastereqn1} at zero
temperature, so to show the method and approximations used, we
evaluate a particular term in \eqn{eqn:mastereqn1} using the
Fourier decomposition of $S(t)$:
\begin{eqnarray}
&&\sum_{k,q} f_\source(\omega_k)(1-f_\drain(\omega_q)) \int_{-\infty}^t\ud \dummy S(t)  \rhoi(\dummy)  S^\dagger(\dummy)\nonumber\\
&&{}=\sum_{mn}\int \ud\omega_k\int \ud\omega_q g_\source g_\drain f_\source(\omega_k)(1-f_\drain(\omega_q))\int_{-\infty}^t\ud \dummy e^{-i(\omega_k-\omega_q+\omega_m)t}e^{i(\omega_k-\omega_q+\omega_n)\dummy}P_m  \rhoi(\dummy)  P_n^\dagger,\nonumber\\
&&{}\approx\pi g_\source g_\drain\sum_{mn}\int \ud\omega_k\int \ud\omega_q  f_\source(\omega_k)(1-f_\drain(\omega_q)) \delta(\omega_k-\omega_q+\omega_n) e^{i(\omega_n-\omega_m)t}P_m  \rhoi(t)  P_n^\dagger,\nonumber\\
&&{}\approx\pi g_\source g_\drain\sum_{n} \int \ud \omega_k f_\source(\omega_k)(1-f_\drain(\omega_k+\omega_n))P_n  \rhoi(t)  P_n^\dagger.\nonumber\\
&&{}=\pi g_\source
g_\drain\sum_n\ramp(\mu_\source-\mu_\drain+\omega_n)P_n  \rhoi(t)
P_n^\dagger,\label{eqn:termintegral}
\end{eqnarray}
where $\ramp(x)=(x+|x|)/2$ is the ramp function and $\mu_i$ is the
chemical potential of lead $i$.
The first equality follows from substituting the Fourier decomposition of $S(t)$, the second equality follows from 
 making the Markov approximation, $\rhoi(\dummy)\rightarrow\rhoi(t)$, the third (approximate) equality we have made a rotating-wave approximation (RWA), where we make the replacement $e^{i(\omega_n-\omega_m)t}\rightarrow\delta_{m,n}$, and finally we integrate over $\omega_k$.  The RWA is reasonable as long as $\splitting\gg\nu^2 e V$, where $\nu=\sqrt{2\pi g_\source g_\drain} \xoo$ is a small quantity.  This is justified when we come to integrate the master equation where terms with $n\neq m$ are rotating sufficiently rapidly to vanish.  Using the same arguments on each term in \eqn{eqn:mastereqn1} results in the general form of the  master equation
\begin{equation}
\dot \rhoi(t)=2\pi g_\source g_\drain\sum_n
\left(\lind[\sqrt{\ramp(e
V+\omega_n)}P_n]\rhoi(t)+\lind[\sqrt{\ramp(-e
V-\omega_n)}P_n^\dagger]\rhoi(t)\right),\label{eqn:mastereqn2}
\end{equation}
where $V=\mu_\source-\mu_\drain$ is the bias applied across the
PC, $\lind[B]\rho=\jump[B]\rho-\A[B]\rho$, $\jump[B]\rho=B\rho
B^\dagger$ and $\A[B]\rho=\frac{1}{2}(B^\dagger B \rho+\rho
B^\dagger B)$.

We may transform the master equation, \eqn{eqn:mastereqn2}, back
to a Schr\"odinger picture to reinstate the free dynamics of the
double well system.  It conveniently turns out that the
transformation of the $P_n$'s back to the Schr\"odinger picture
merely multiplies them by the unitary factor $e^{i \omega_n t}$.
The Lindblad superoperators are invariant under such a
transformation, so in the Schr\"odinger picture
\begin{equation}
\dot \rhos(t)=-i\commute{\hsys}{\rhos(t)}+\sum_n \left(\U(e
V+\omega_n)\lind[c_n]\rhos(t) + \U(-e
V-\omega_n)\lind[c_n^\dagger]\rhos(t)\right)\equiv \Lind
\rhos(t),\label{eqn:mastereqnSP}
\end{equation}
where $\U(x)$ is the unit step function.  For convenience,  we
have defined $c_n=\sqrt{\ramp(e V+\omega_n)}P_n$, so that
\begin{eqnarray}
c_1&=&\nu \sqrt{e V+\splitting}\sin(\theta)\op{g}{e},\label{eqn:c1}\\
c_2&=&\nu \sqrt{|e V-\splitting|}\sin(\theta)\op{e}{g},\label{eqn:c2}\\
c_3&=&
\nu \sqrt{e V}\cos(\theta)\sigma_z^{(e)}+\Tnd\sqrt{e
V}.\label{eqn:c3}
\end{eqnarray}
We will show later, using a explicit model of the detection
process, that $\Tnd=\sqrt{2\pi g_\source g_\drain}\too$.

\section{Analytic solutions of the unconditional master equation}

\subsection{High bias regime}

In the high bias regime,  $\U(-e V-\omega_n)=0$ for all $n$, and
the interaction picture master equation takes the form
\begin{equation}
\dot \rho_I(t)= \sum_n \lind[c_n] \rho_I(t) 
.\label{eqn:mastereqnHighBias}
\end{equation}
We can solve the unconditional master equation in this regime
exactly. We present the solution in the interaction picture, since
it contains the relevant decay time scales. (It is straightforward
to find the corresponding dynamics in the Schr\"odinger picture
via the transformation $\rho(t)= e^{-i \hsys t}\rhoi(t) e^{i \hsys
t}$). The general solution is given by
\begin{eqnarray}
\rho_{ge}(t)&=&e^{-{\Gamma_d} (1+\cos^2(\theta))t/2}\rho_{ge}(0),\nonumber\\
\rho_{gg}(t)&=& \frac{\splitting+e V}{2 e V} -
e^{-\Gamma_d\sin^2(\theta)t}(\frac{\splitting+e V}{2 e
V}-\rho_{gg}(0)),\label{eqn:solnHB}
\end{eqnarray}
where we have defined $\Gamma_d=2\nu^2 e V$, consistent with
\cite{goa01a}.

\subsection{Low bias regime}

Our analysis is also valid in the low bias regime, $\splitting>e
V$. The unconditional master equation in this case, in the
interaction picture, may be written
\begin{equation}
\dot \rho_I(t)=
\lind[c_1]\rho_I(t)+\lind[c_2^\dagger]\rho_I(t)+\lind[c_3]\rho_I(t)
.\label{eqn:mastereqnLowBias}
\end{equation}
Again, we can solve the unconditional master equation in this
regime exactly, and we give the solution in the interaction
picture,
\begin{eqnarray}
\rho_{ge}(t)&=&e^{-(\Gamma_d\cos^2(\theta)+\nu^2\splitting\sin^2(\theta))t}\rho_{ge}(0) ,\nonumber\\
\rho_{gg}(t)&=& 1 -e^{-2
\nu^2\splitting\sin^2(\theta)t}(1-\rho_{gg}(0)).\label{eqn:solnLB}
\end{eqnarray}
Note that at $\splitting=e V$, \eqns{eqn:solnHB} and
(\ref{eqn:solnLB}) agree.

\section{Conditional master equations}
\label{sec:CME}

It is also possible to describe the dynamics of the continuously
measured charge qubit using a \emph{conditional} master equation
(CME). This describes the evolution of the state of the system,
conditioned on the previous detector output. The CME is useful,
since it allows individual measurement runs to be simulated, and
also allows a straightforward calculation of detector output
statistics, such as the average current and the power spectrum of
current fluctuations. In this section, we derive a CME using an
explicit model of the measurement process in terms of projective
measurements of the number of electrons that have tunnelled
through the PC. Our derivation follows a similar argument as that
presented in \cite{brepet} for the case of a
continuously observed fluorescing atom.

Owing to the different types of allowed jump processes in each
regime, we treat the high and low bias regimes separately.

\subsection{High bias regime}

In order to derive the CME in the high bias ($e V > \splitting$)
regime, we first consider the unconditional evolution of the qubit
over a short time interval $\delta t$, which is taken to be
sufficiently short that no more than one electron tunnels across
the PC during this interval. This unconditional evolution can be
written as a sum of two conditional terms, as
\begin{equation}
\rho_I(t_0 + \delta t) = \tilde{\rho}_{0c}(t_0 + \delta t) +
\tilde{\rho}_{1c} (t_0 + \delta t) \,. \label{eqn:uncondVScond1HB}
\end{equation}
Here, $\tilde{\rho}_{0c}$ is the unnormalized conditional density
matrix for the qubit, corresponding to having observed zero
electrons pass through the PC in the interval $\delta t$, whereas
$\tilde{\rho}_{1c}$ is the unnormalized conditional density matrix
for the qubit, corresponding to having observed a single electron
pass through the PC, in the source to drain direction, in the same
interval. In the high bias regime, tunnelling in the reverse
direction (drain to source) does not conserve energy and therefore
is strongly suppressed on timescales of interest, as we discuss
further below. The normalization of these conditional states is
chosen such that $\mathrm{Tr} \left \{ \tilde{\rho}_{0(1)c}(t_0 +
\delta t)\right \}$ is the probability that zero (one) electrons
tunnelled through the PC in the interval $\delta t$.

Assuming that at time $t_0$, the total density matrix for the
system and detector may be written in the factorized form
$\rho(t_0) \otimes \vert 0 \rangle \langle 0\vert$, where $\vert 0
\rangle \langle 0\vert$ represents the zero temperature state of
the source and drain leads before any electrons have tunnelled,
$\tilde{\rho}_{1c} (t_0 + \delta t)$ may be written
\begin{equation}
\tilde{\rho}_{1c} (t_0 + \delta t) = \mathrm{Tr}_{S,D} \left\{
\Pi_1 U_I(t_0,t_0+ \delta t) \left(\rho(t_0) \otimes \vert 0
\rangle \langle 0\vert \right) U^\dag_I(t_0,t_0+ \delta t) \Pi_1
\right\} \label{eqn:rho1conditional1HB}
 \,.
\end{equation}
Here,
\begin{equation}
\Pi_1 = \sum_{\alpha>k_S^F, \beta<k_D^F} a^\dag_{D,\alpha}a_{S,\beta} \vert 0
\rangle \langle 0\vert a^\dag_{S,\beta}a_{D,\alpha}
\end{equation}
is the projector onto the subspace where only one electron has
tunnelled through the PC in the source to drain direction,
$U_I(t_0,t_0+ \delta t)$ is the interaction picture time evolution
operator for the interval $\delta t$ and $k^F_{S(D)}$ is the Fermi wavenumber for the source (drain) lead.  For brevity we will suppress the range of the summation below. Note that Eq.
(\ref{eqn:rho1conditional1HB}) follows directly from the
projection postulate of quantum mechanics.

An approximate expression for $\tilde{\rho}_{1c} (t_0 + \delta t)$
can be found by first expanding the time evolution operator up to
second order in $\delta t$, i.e.
\begin{equation}
U_I(t_0,t_0+ \delta t) \approx 1 - i \int_{t_0}^{t_0+\delta t} dt'
H_I(t') - \int_{t_0}^{t_0+\delta t} dt' \int_{t_0}^{t'} dt''
H_I(t') H_I(t'') \,,
\end{equation}
and then substituting into Eq. (\ref{eqn:rho1conditional1HB}) to
give
\begin{eqnarray}
\tilde{\rho}_{1c} (t_0 + \delta t) & = & \sum_{\alpha,\beta}
\langle 0\vert a^\dag_{S,\beta}a_{D,\alpha} U_I(t_0,t_0+ \delta t)
\vert 0 \rangle \rho(t_0) \langle 0\vert
 U^\dag_I(t_0,t_0+ \delta t) a^\dag_{D,\alpha} a_{S,\beta}  \vert 0 \rangle,
\\
& \approx &  \sum_{\alpha,\beta} \int_{t_0}^{t_0+\delta t}dt' 
\int_{t_0}^{t_0+\delta t} dt'' \langle 0\vert
a^\dag_{S,\beta}a_{D,\alpha} H_I(t') \vert 0 \rangle \rho(t_0)
\langle 0\vert
 H_I(t'') a^\dag_{D,\alpha} a_{S,\beta}  \vert 0 \rangle. \label{eqn:rho1conditional2HB}
\end{eqnarray}
Note that the zeroth and second order terms in the expansion of
$U_I(t_0,t_0+ \delta t)$ do not contribute to the above expression
for $\tilde{\rho}_{1c} (t_0 + \delta t)$, since the expression
contains a projector onto the subspace where only one electron has
tunnelled through the PC.

Evaluating the integrals in Eq. (\ref{eqn:rho1conditional2HB})
leads to the expression
\begin{equation}
\tilde{\rho}_{1c} (t_0 + \delta t) =  \sum_{\alpha,\beta} \left[
{f}_{\alpha \beta 1} P_1 + {f}_{\alpha \beta 2} P_2 + {f}_{\alpha
\beta 3} P_3\right] \rho(t_0) \left[ {f}_{\alpha \beta 1} P_1 +
{f}_{\alpha \beta 2} P_2 + {f}_{\alpha \beta 3} P_3\right]^\dag
\,, \label{eqn:rho1conditional3HB}
\end{equation}
where $P_i$ are the operators introduced in Sec.
\ref{sec:UME}, and
\begin{eqnarray}
{f}_{\alpha \beta 1} &=& \delta t \,
e^{-i(\omega_{\beta}-\omega_{\alpha}+\splitting)(t_0 + \delta t /
2)}
\mathrm{sinc} \left( (\omega_{\beta}-\omega_{\alpha}+\splitting) \delta t /2\right) H(\mu_S - \omega_\alpha) H(\omega_\beta-\mu_D ) \,, \\
{f}_{\alpha \beta 2} &=& \delta t \,
e^{-i(\omega_{\beta}-\omega_{\alpha}-\splitting)(t_0 + \delta t /
2)}
\mathrm{sinc} \left( (\omega_{\beta}-\omega_{\alpha}-\splitting) \delta t /2\right) H(\mu_S - \omega_\alpha) H(\omega_\beta-\mu_D ) \,, \\
{f}_{\alpha \beta 3} &=& \delta t \,
e^{-i(\omega_{\beta}-\omega_{\alpha})(t_0 + \delta t / 2)}
\mathrm{sinc} \left( (\omega_{\beta}-\omega_{\alpha}) \delta t
/2\right) H(\mu_S - \omega_\alpha) H(\omega_\beta-\mu_D ) \,.
\end{eqnarray}
For values of $\delta t \gg \splitting^{-1}$, the cross terms in
Eq. (\ref{eqn:rho1conditional3HB}) make a very small contribution,
and can be neglected. The remaining terms in Eq.
(\ref{eqn:rho1conditional3HB}) may be evaluated by converting the
sums into integrals via the replacement $\sum_{k}\rightarrow\int
\ud\omega_{k} g_{\source(\drain)}(\omega_{k})$ as in Sec.
\ref{sec:UME}. Finally, again for $\delta t \gg \splitting^{-1}$, we make the following replacement in the integrands
\begin{equation}
{\sin^2( x \,\delta t /2)}/ {x^2}\rightarrow
\pi \delta t \, \delta(x)/4 \,. \label{eqn:deltafunc}
\end{equation}
Thus Eq.\ (\ref{eqn:rho1conditional3HB}) reduces to
\begin{equation}
\tilde{\rho}_{1c} (t_0 + \delta t) = \delta t \sum_{n=1}^3
\jump[c_n] \rho (t_0) \,, \label{eqn:rho1conditional4HB}
\end{equation}
where the jump operators $\{c_n\}$ are given by Eqs. (\ref{eqn:c1}
- \ref{eqn:c3}), with $\Tnd=\sqrt{2\pi g_\source g_\drain}\too$.

Note that using the replacement of Eq. (\ref{eqn:deltafunc})
amounts to a statement of energy conservation (on sufficiently
long timescales). For this reason, processes such as reverse
(drain to source) tunnelling are suppressed in the high bias
regime, since they do not conserve energy.

The unnormalized state conditioned on having observed no
tunnelling events in the time interval $\delta t$,
$\tilde{\rho}_{0c}$, can be found by repeating the above
derivation, but replacing $\Pi_1$ in Eq.
(\ref{eqn:rho1conditional1HB}) with $\Pi_0 = \vert 0 \rangle
\langle 0 \vert$. However, it is also possible to find
$\tilde{\rho}_{0c}$ by a more direct route, by noting that the
normalized, unconditional state satisfies
\begin{equation}
\rho_I(t_0 + \delta t) = \left(1 + \delta t \sum_{n=1}^3
\lind[c_n]
 \right) \rho_I(t_0) \,. \label{eqn:uncondVScond2HB}
\end{equation}
Combining Eqs. (\ref{eqn:uncondVScond1HB}),
(\ref{eqn:rho1conditional4HB}) and (\ref{eqn:uncondVScond2HB}),
gives
\begin{equation}
\tilde{\rho}_{0c} (t_0 + \delta t) = \left(1 - \delta t
\sum_{n=1}^3 \A[c_n] \right) \rho_I(t_0) \,.
\label{eqn:rho0conditional1HB}
\end{equation}

Equations (\ref{eqn:rho1conditional4HB}) and
(\ref{eqn:rho0conditional1HB}) can be combined into a single,
normalized, CME for the conditional state of the system, in the
Schr\"odinger picture, as
\begin{equation}
d \rho_c(t) = dN(t) \left( \frac{\sum_{n = 1}^3 \jump[c_n]
}{P_{1c}(t)} - 1\right) \rho_c(t) + dt \left(-\sum_{n = 1}^3
\A[c_n] \rho_c(t) + P_{1c}(t) \rho_c(t) -
i[\hsys,\rho_c(t)]\right)\,. \label{eqn:CMEHB}
\end{equation}
Here, $\rho_c(t)$ denotes the normalized conditional state of the
system at time $t$, and $P_{1c}(t) = \mathrm{Tr} \left\{\sum_{n
= 1}^3 \jump[c_n] \rho_c(t) \right\}$ is the probability density
of observing a single electron tunnelling event. We have also introduced the classical stochastic increment
$dN(t)\in\{0,1\}$ which denotes the number of electron tunnelling
events in the interval $dt$. The expectation value of this
increment is given by
\begin{equation}
E[dN(t)] = P_{1c}(t)dt =\mathrm{Tr} \left\{ \sum_{n=1}^3
\jump[c_n] \rho_c(t) \right\} d t\,.
\end{equation}

Note that although the CME of Eq.\ (\ref{eqn:CMEHB}) has been
written in terms of an infinitesimal time increment $dt$, it
should be understood that in deriving the CME we have made use of
a \emph{coarse-graining} in time. In particular, neglecting the
cross terms in Eq. (\ref{eqn:rho1conditional3HB}) means that the
derived CME neglects dynamics on fast timescales of order
$\phi^{-1}$. Implications of this will be discussed further in
Sec. \ref{sec:comparison}.

\subsection{Low bias regime}

In the low bias ($eV < \splitting$) regime, the CME must also take
into account tunnelling in the reverse (drain to source)
direction. Note that these reverse tunnelling events are, in
principle, distinguishable from the forward tunnelling events.
Thus the unconditional evolution over the interval $\delta t$ may
be written as the sum of \emph{three} conditional terms:
\begin{equation}
\rho_I(t_0 + \delta t) = \tilde{\rho}_{0c}(t_0 + \delta t) +
\tilde{\rho}_{1^+c} (t_0 + \delta t) + \tilde{\rho}_{1^-c} (t_0 +
\delta t)\,, \label{eqn:uncondVScond1LB}
\end{equation}
where $\tilde{\rho}_{1^+c} $ ($\tilde{\rho}_{1^-c} $) denotes the
unnormalized conditional state of the qubit corresponding to
having observed a single tunnelling event in the forward (reverse)
direction in the interval $\delta t$, and $\tilde{\rho}_{0c}$
corresponds to having observed no tunnelling events in the same
interval.

The derivation of the CME now proceeds in the same manner as in
the high bias case, except that now three projectors are used,
corresponding to the different distinguishable measurement
outcomes:
\begin{eqnarray}
\Pi_0 &=& \vert 0
\rangle \langle 0\vert \,, \\
\Pi_{1^+} &=& \sum_{\alpha, \beta} a^\dag_{D,\alpha}a_{S,\beta}
\vert 0
\rangle \langle 0\vert a^\dag_{S,\beta}a_{D,\alpha} \,, \\
\Pi_{1^-} &=& \sum_{\alpha, \beta} a^\dag_{S,\alpha}a_{D,\beta}
\vert 0 \rangle \langle 0\vert a^\dag_{D,\beta}a_{S,\alpha} \,.
\end{eqnarray}

The resulting CME may be written
\begin{multline}
d \rho_c(t) = dN^+(t) \left( \frac{\jump[c_1] + \jump[c_3]
}{P_{1^+c}(t)} - 1\right) \rho_c(t) + dN^-(t) \left(
\frac{\jump[c_2^\dag]
}{P_{1^-c}(t)} - 1\right) \rho_c(t)\\
 + dt \left(-\left\{\A[c_1] + \A[c_2^\dag] + \A[c_3]\right\}
 \rho_c(t) + \left[P_{1^+c}(t) +
P_{1^-c}(t)\right]\rho_c(t) - i[\hsys,\rho_c(t)]\right)\,.
\label{eqn:CMELB}
\end{multline}
Here, $dN^\pm(t)\in\{0,1\}$ denotes the number of forward
(reverse) electron tunnelling events in the interval $dt$. The
expectation values of these increments are given by
\begin{eqnarray}
E[dN^+(t)] &=& P_{1^+c}(t)dt =\mathrm{Tr} \left\{ \jump[c_1] \rho_c(t) + \jump[c_3] \rho_c(t) \right\} d t\,, \\
E[dN^-(t)] &=& P_{1^-c}(t)dt =\mathrm{Tr} \left\{ \jump[c_2^\dag]
\rho_c(t) \right\} d t\,.
\end{eqnarray}

\section{Steady state results} \label{sec:SteadyState}
\subsection{Pseudo-thermal ground state
occupation}\label{sec:pstherm}

We can use the unconditional master equation in each regime to
compute the unconditional, steady-state probability of the qubit
to be found in its ground state.  It is given by
$\rho_{gg}(\infty)=\bra{g}\rho(\infty)\ket{g}$.  From
\eqns{eqn:solnHB} and (\ref{eqn:solnLB}), we see that
$\rho_{gg}(\infty)$  is
\begin{equation}
\rho_{gg}(\infty)=\left\{\begin{array}{ c c}
     \frac{e V+\splitting}{2e V} & \textrm{ if }\splitting<e V \\
     1 & \textrm{ if }\splitting>e V
\end{array}\right.
\end{equation}

\begin{figure}
\includegraphics[width=8cm]{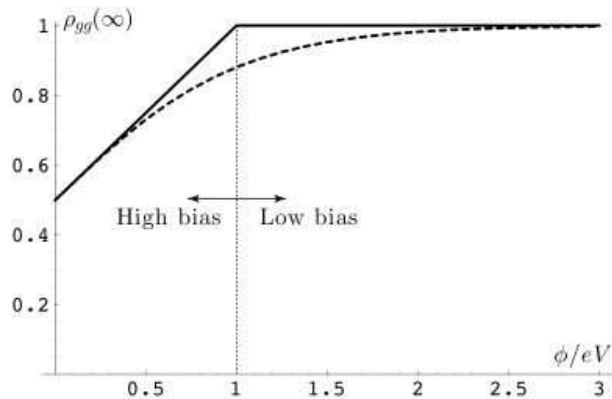}
\caption{\label{fig:QuasiThermal} Equilibrium ground state
occupation probability for a qubit near  a PC (solid), and the
thermal equilibrium ground state occupation probability,
$\rho_{gg}^\mathrm{therm}(\infty)$ for a qubit in contact with a
heat bath at temperature $T=e V/2$ (dashed).}
\end{figure}

In contrast, the ground state occupation of a qubit in thermal
equilibrium with a heat bath at temperature $T$ is given by
$\rho_{gg}^\mathrm{therm}(\infty)=\frac{e^{\splitting/\Boltz
T}}{1+e^{\splitting/ \Boltz T}}$.  Figure \ref{fig:QuasiThermal}
shows these two ground state probabilities for the case where
$\Boltz T=e V/2$, and there is an evident analogy between the PC
bias voltage and an external heat bath, even though the leads are
each nominally at zero temperature, that is
$f_{\source,\drain}(\omega)=\U(\mu_{\source,\drain}-\omega)$. As
the PC bias gets larger the relative fraction of each component of
the mixture tends to $1/2$, analogously to a qubit in contact with
a heat bath at very high temperature.

It has been suggested elsewhere that the PC bias acts somewhat
like a heat bath \cite{moz02}, and we note that this
correspondence is a natural conclusion of our analysis.  It may be
seen that even in the limit $e V\rightarrow0$, the PC still causes
the qubit to decay to its ground state. We can see from
\eqn{eqn:solnLB} that in this limit, the energy relaxation time is
given by $\trel^{-1}=2  \nu^2\splitting\sin^2(\theta)$, and the
dephasing time is $\tdeph^{-1}= \Gamma_d\cos^2(\theta)+\nu^2\splitting\sin^2(\theta)$.
When the PC bias voltage is zero, we see that $\tdeph=2\trel$,
indicating that the dephasing is solely due to energy relaxation.
This zero-bias dephasing arising from relaxation is the solid
state analogue of the optical decay of an atom in a vacuum.

This is a significant practical issue, since it means that the PC
cannot be turned off merely by making the PC bias zero. To perform
coherent quantum logic operations on the qubit in the presence of
the PC, with error rates small enough to permit fault tolerant
quantum computation, will require an extra gate to control the
coupling, $\nu$, between the qubit and the PC. $\nu$ can be
significantly reduced by raising the height of the tunnel barrier
between the source and drain leads of the PC. Then when the PC is
off (i.e.\ $e V=0$), $\nu^2 \splitting$ can be made much smaller
than $\splitting$ by a factor of order the fault tolerant
threshold \cite{pre98a} $\sim10^{-4}$, by raising this barrier.

\subsection{Steady state current}
\label{sec:Iss}

The steady state current can be calculated using expectation
values of the stochastic increments defined in Sec.\ \ref{sec:CME}.
In the high bias regime, the steady state current is given by
$I_{ss} = e E[dN(t)/dt]_{t \to \infty} = e \Tr \{ \sum_{n=1}^3
\jump[c_n] \rho(\infty)\}$. In the low bias regime, it is given by
$I_{ss} = e E[(dN^+(t) - dN^-(t))/dt]_{t \to \infty} = e \Tr \{
\jump[c_1]\rho(\infty) + \jump[c_3]\rho(\infty) -
\jump[c_2^\dag]\rho(\infty)\}$.

The current can be conveniently expressed in terms of the expected
currents, $\current_l$ and $\current_r$, for the case when the
tunnelling between dots is switched off ($\theta = 0$), and the
qubit is in the localized state $\vert l \rangle \langle l \vert$
or $\vert r \rangle \langle r \vert$, respectively. Then the
average value of these currents is given by
$\iloc=\frac{\current_r+\current_l}{2}=e(\Tnd^2+\nu^2) e V$ and
the difference is
$\diloc=\current_r-\current_l=-4 e\nu\Tnd e V$.



To first order in $\diloc$, we find
\begin{equation} \label{eqn:Iss}
I_{ss}=\left\{\begin{array}{ c c}
     \iloc- \frac{ \cos(\theta)\splitting}{2 e V}\diloc=\iloc- \frac{\epsilon}{2 e V}\diloc & \textrm{ if }\splitting<e V \\
     \iloc- \frac{\cos(\theta)}{2}\diloc=\iloc- \frac{\epsilon}{2\splitting} \diloc& \textrm{ if }\splitting>e V
\end{array}\right.
.
\end{equation}
Using $I_{ss}$ we can compute the DC conductance
\begin{equation}
G =I_{ss}/V=\left\{\begin{array}{ l c}
     \Tnd^2+\nu^2+2\nu\Tnd\epsilon/(e V)& \textrm{ if }\splitting<e V, \\
     \Tnd^2+\nu^2+2\nu\Tnd\epsilon/\splitting & \textrm{ if }\splitting>e V,
\end{array}\right. \label{eqn:conductance}
\end{equation}
in units of $e^2/h$. Note that these results agree with those
found in Ref. \onlinecite{eng03}, and given implicitly in the shot
noise component of the power spectra of Ref. \onlinecite{stace:136802}.

\begin{figure}
\includegraphics[width=7cm]{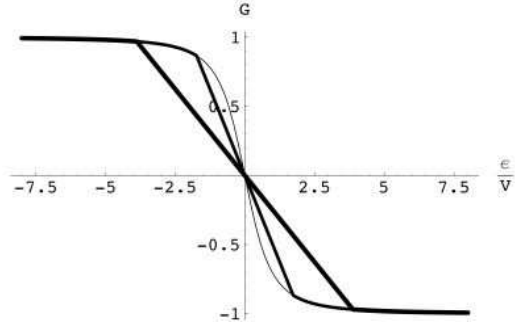}
\caption{\label{fig:SSconductance} The steady state conductance,
$G$, versus $\epsilon/eV$, for different values of the double well
tunnelling rate, (light) $eV/\Delta=0$, (medium) $eV/\Delta=2$ and
(dark) $eV/\Delta=4$}
\end{figure}

Equation (\ref{eqn:Iss}) indicates that the steady state current
depends on the steady state charge distribution of the qubit,
which in turn depends on the qubit bias $\epsilon$.  As $\epsilon$
varies the conductance changes accordingly, which is qualitatively
in agreement with observations of charge delocalization made in
recent experiments \cite{gar03,elz03,dicarlo:226801,rushforth:113309}.
This behaviour is shown in \fig{fig:SSconductance}.  Note that the
cusp at $\epsilon=\sqrt{(eV)^2-\Delta^2}$ appears when
$eV>\Delta$.

This result suggests an improved method for performing
spectroscopy of the qubit using steady state conductance
measurements through the PC: the value of the applied voltage,
$eV$, at which a cusp in the conductance (when plotted as a
function of $\epsilon$) becomes evident \emph{directly} gives the
unbiased qubit tunnelling rate, $\Delta$. Alternatively, observing
the conductance as a function of $eV$, for a fixed value of
$\epsilon$, yields the splitting $\splitting$, by determining the
voltage at which the conductance changes from low- to high-bias
behaviour in Eq. (\ref{eqn:conductance}). These methods offer a
potential improvement over existing, indirect techniques
\cite{dicarlo:226801}, in which an estimate of $\Delta$ is extracted
from a fit to a model, which in turn is limited by the uncertainty
in the estimated electron temperature.

\section{Power spectra of current correlations}

Using the conditional master equations, we are able to evaluate
the steady state power spectrum of the two time current
correlation function
\begin{equation}
G(\tau)=E[\current(t+\tau)\current(t)]-E[\current(t+\tau)]E[\current(t)],
\end{equation}
and we adopt the same procedure as described in the appendix of
\citet{goa01a}.  The power spectrum is the Fourier transform of
this quantity,
\begin{equation}
S(\omega)=2\int_{-\infty}^\infty d\tau\, G(\tau)e^{-i\omega
\tau}=\int_0^\infty d\tau\, G(\tau)\cos(\omega \tau),
\end{equation}
since $G(\tau)=G(-\tau)$.

\subsection{High bias power spectrum}

To compute the steady state correlation function, we take the
limit $t\rightarrow\infty$, so $\rho(t)\rightarrow\rho_\infty$.
Using the relation $I(t) \,dt=e\, dN(t)$, where $dN(t)\in\{0,1\}$
is the number of electron tunnelling events in the time interval
$(t, t+dt)$ and,
\begin{eqnarray}
E[dN(t+\tau)dN(t)]&=&\prob[dN(t)=1]E[dN_c(t+\tau)|_{dN(t)=1}],\nonumber\\
\prob[dN(t)=1]&=& dt \tr{\tilde \rho_{1c}(t+dt)},\nonumber\\
E[dN_c(t+\tau)|_{dN(t)=1}]&=& dt \Tr\{\sum_n \jump[c_n]E[\tilde \rho_1c(t+\tau)|_{dN(t)=1}]\},\nonumber\\
E[\tilde \rho_{1c}(t+\tau)|_{dN(t)=1}]&=&e^{\Lind\,\tau} \tilde \rho_{1c}(t+dt)/\tr{\tilde \rho_{1c}(t+dt)},\nonumber\\
dN(t)^2 &=& dN(t),\label{eqn:expectationidentities}
\end{eqnarray}
 we obtain the correlation function in the high bias regime for $\tau>dt$
\begin{equation}
\ghb(\tau)=\left\{\begin{array}{ l l}
      e^2(\Tr\{\sum_{n,n'}\jump[c_n]e^{\Lind\,\tau}\jump[c_{n'}]\rho_\infty\}-\Tr\{\sum_{n}\jump[c_n]\rho_\infty\}^2) &  \textrm{ for }\tau>dt,  \\
      e^2 \Tr\{\sum_{n}\jump[c_n]\rho_\infty\} \delta(\tau)&   \textrm{ for }\tau=dt.
\end{array}\right.
\end{equation}
We write $\ghb(\tau)$ in terms of $\iloc$ and $\diloc$:
\begin{equation}
\ghb(\tau)=e( \iloc-\diloc\frac{\splitting}{2e
V}\cos(\theta))\delta(\tau)+e^{-\Gamma_d {\sin^2 (\theta
)}\tau}e^2(1-\frac{\splitting^2}{(e V)^2})\nu^2(2\Tnd e
V\cos(\theta)+\nu \splitting \sin^2(\theta))^2.
\end{equation}
Dropping the small term $\nu \splitting \sin^2(\theta)$, which is
formally equivalent to keeping only the lowest order term in a
series expansion in $\diloc$, and taking the Fourier transform
gives
\begin{equation}
\shb(\omega)=S_0-e\diloc\frac{\splitting}{V}\cos(\theta)+\frac{\diloc^2\Gamma_d(1-\frac{\splitting^2}{(e
V)^2})\sin^2(2\theta)}{4(\Gamma_d^2\sin^4(\theta)+
\omega^2)}\label{eqn:PSHB}
\end{equation}
where $S_0=2e\iloc$.

\subsection{Low bias power spectrum}

\begin{figure*}
\includegraphics[height=6cm]{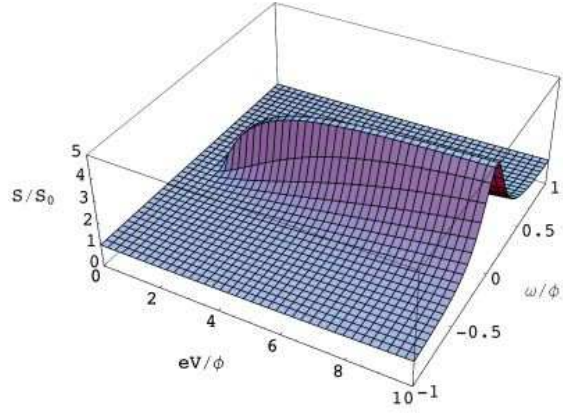}
\caption{\label{fig:PowerSpectrum}  Power spectra, $S(\omega)/S_0$, computed in this paper using \eqns{eqn:PSHB} and (\ref{eqn:PSLB}),  as a function of $eV/\splitting$ for $\theta=\pi/4$, $\nu^2=0.02$. 
}
\end{figure*}

The correlation function in the low bias regime turns out to be
time independent.  In this regime, there are two distinguishable
jump processes that can occur: source to drain electron tunnelling
(both elastic and inelastic ), and drain to source tunnelling
(inelastic only).  Therefore the current is related to the number
of jumps in the time interval by $I(t)\,dt=e(dN^+(t)-dN^-(t))$,
where $dN^{+(-)}(t)$ counts the number of source-to-drain
(drain-to-source) tunnelling events in the time interval
$(t,t+dt)$.

To evaluate the expectation values in $G(\tau)$, we use slightly
generalized versions  of \eqns{eqn:expectationidentities}, for
instance
\begin{eqnarray}
\prob[dN^{\pm}(t)=1]&=& dt\tr{\tilde \rho_{1^\pm c}(t+dt)},\nonumber\\
 E[dN_c^x(t+\tau)|_{dN^y(t)=1}]&=& dt\Tr\{\jump^x E[\tilde \rho_1c(t+\tau)|_{dN^y(t)=1}]\}\textrm{ for } x,y=\pm.\nonumber
 \end{eqnarray}
 As before, to compute the steady state power spectrum, we make the replacement $\rho(t)\rightarrow\rho_\infty=\op{g}{g}$, for the low bias regime.  It is evident that $\jump^-\rho_\infty=\jump[c_2^\dagger]\rho_\infty=\jump[c_1]\rho_\infty=0$, so we find that
\begin{eqnarray}
\glb(\tau)&=&\left\{\begin{array}{ l l}
      e^2(\Tr\{\jump[c_3]e^{\Lind\,\tau}\jump[c_{3}]\rho_\infty\}-\Tr\{\jump[c_3]\rho_\infty\}^2) &  \textrm{ for }\tau>dt,  \\
      e^2 \Tr\{\jump[c_3]\rho_\infty\} \delta(\tau)&   \textrm{ for }\tau=dt.
\end{array}\right.\\
&=&e^2(\Tnd+ \nu\cos(\theta))^2 e V\delta(\tau)
\end{eqnarray}
Thus, the steady state low bias current correlation function is
just due to elastic tunnelling through the PC.  The low bias power
spectrum is
\begin{equation}
\slb(\omega)=S_0-e\, \diloc
\cos(\theta)+O(\diloc^2).\label{eqn:PSLB}
\end{equation}

The power spectra computed in this paper are shown in
\fig{fig:PowerSpectrum} for $\nu^2=0.02$.  Note the sharp change
at $eV=\phi$.

\subsection{Comparison with previous results}
\label{sec:comparison}

\begin{figure}
\subfigure[]{\includegraphics[height=5cm]{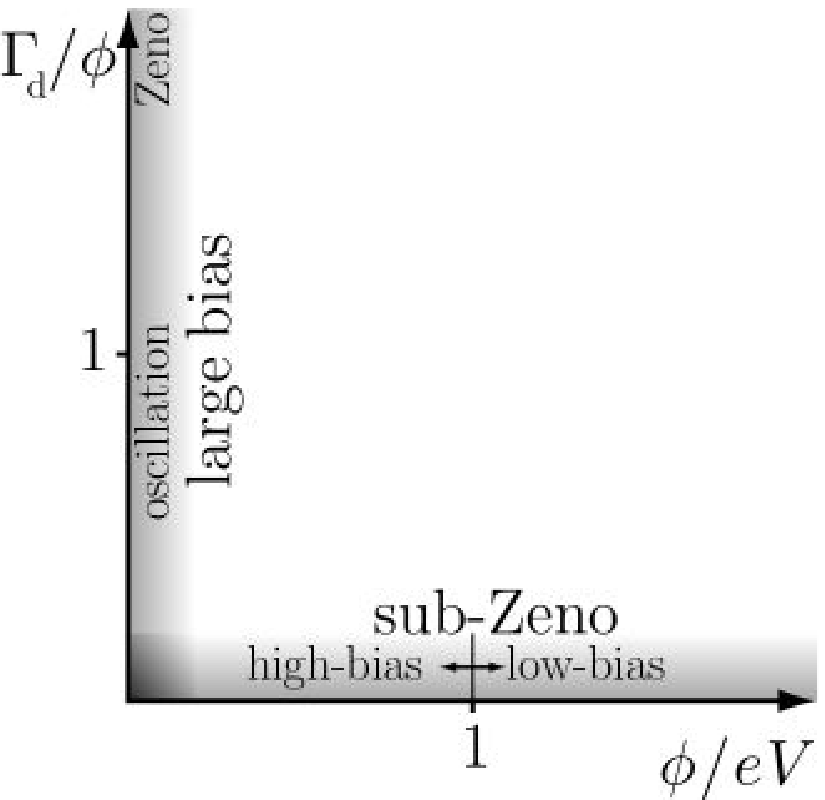}\label{fig:RegionOfValidity}}\hfil
\subfigure[]{\includegraphics[height=5cm]{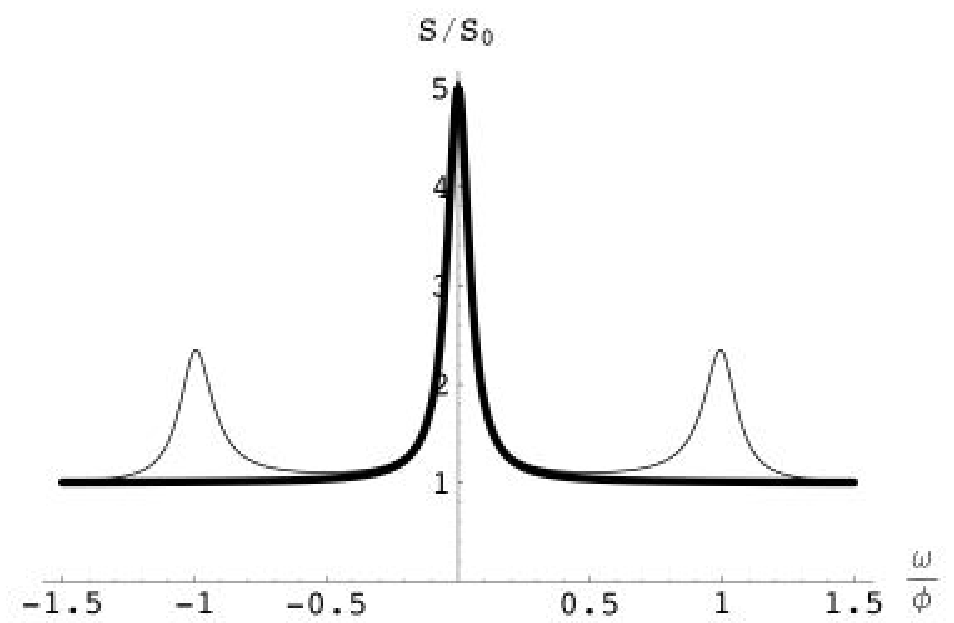}\label{fig:comparison}}
\caption{ (a) Phenomena predicted in this paper (sub-Zeno) and
previous work (large bias). (b) The power spectrum computed in
this paper with $\Gamma_d=0.1$, $\theta=\pi/4$ and $e V=100\phi$
(dark), and the power spectrum computed by \cite{goa01a}, (see
appendix \ref{app:GMPS}), with the same parameters (light).}
\end{figure}

The work presented here is valid when $\Gamma_d\ll\phi$, whilst
previous work  has dealt with the large bias limit, $\phi\ll eV$
\cite{goa01a,kor01a,kor01b}.  These limits are shown schematically
in \fig{fig:RegionOfValidity}.  The limit $\Gamma_d\ll\phi\ll eV$
is common to both this work and previous analyses
\cite{goa01a,kor01a,kor01b}, so we can compare the predictions of
each in this limit.

The current correlation function,  $\ggm(\tau)$, given in
\cite{goa01a} (subscript GM denotes the authors initials) is only
valid for $\epsilon=0$, but it is straightforward to calculate the
general result, which is useful in order to compare our results
with previous work.  We present the general calculation in
Appendix \ref{app:GMPS}.  Figure \ref{fig:comparison} shows the
power spectra computed in this paper (dark) and using $\ggm(\tau)$
(light) for large bias, $\phi/e V=0.01$, and weak coupling,
$\Gamma_d=0.1$.  The central peak at $\omega=0$, due to inelastic
transitions between the qubit energy eigenstates, shows good
agreement between the results. There is a marked difference
between the spectra at frequencies near $\pm\phi$.  In particular
the peaks predicted in \cite{goa01a} are absent from our results.
However, this difference can be understood by noting that, in
deriving our CME in Sec. \ref{sec:CME}, we have employed a
temporal `coarse graining', which implies that our CME does not
describe dynamics on short timescales of order $\phi^{-1}$.
Therefore, our expressions for the power spectra (which are
derived from our CME) are valid only for frequencies $\omega \ll
\phi$.

In the low bias case, we predict a completely flat power spectrum
(Eq. (\ref{eqn:PSLB})). Although this prediction is also based on
a coarse-grained CME, there is reason to expect that this should
be valid at all frequencies. In particular, since the qubit
rapidly relaxes to the (pure) ground state $\vert g \rangle$,
which is a stationary state, there should be no oscillatory signal
in the PC current, and no peaks in $\slb(\omega)$ should be seen
at $\omega = \phi$. This prediction is in agreement with those
made independently in Ref. \cite{shn02}.

\section{Measurement Time}

The detector partially projects the qubit onto the energy
eigenbasis, as seen in the form of jump operator $c_3$.  The
energy eigenbasis is therefore the `preferred' basis for the
detector.

An important quantity of interest is the measurement time.  This
is the time taken for the measurement to project a qubit initially
prepared in an equal superposition of energy eigenstates onto one
or other of them. We therefore take
$\rho(0)=(\ket{e}+\ket{g})(\bra{e}+\bra{g})/2$.  Following
\citet{goa01}, we compute the rate of change of
$E[dz_c^2(0)]\equiv E[\Tr\{\sigma_z^{(e)}d\rho_c(0)\}^2]$.

We present the calculation for high bias, and note that we obtain
the same result for low bias.  Using the fact that
\begin{equation}
d\rho_c(t)=\rho_c(t+dt)-\rho_c(t)=dN(t)\rho_{1c}(t+dt)+(1-dN(t))\rho_{0c}(t+dt)-\rho_c(t),
\end{equation}
$dN(t)^2=dN(t)$ and $\Tr\{\sigma_z^{(e)}\rho(0)\}=0$, we find that
to first order in $dt$
\begin{eqnarray}
E[dz_c^2(0)]&=&E[dN(0) \Tr\{\sigma_z^{(e)}\rho_{1c}(dt)\}^2]/2,\nonumber\\
&=&\frac{\Tr\{\sigma_z^{(e)}\tilde\rho_{1c}(dt)\}^2}{2\Tr\{\tilde\rho_{1c}(dt)\}},\nonumber\\
&=&\Gamma_d\cos^2(\theta) dt+O(\nu^3\phi).\label{eqn:dz2}
\end{eqnarray}
Thus the measurement time is
$\tmeas^{-1}=\Gamma_d\cos^2(\theta)+O(\nu^3\phi)$.  We note that
$\tmeas\geq\tdeph$, where $\tdeph$ is the dephasing time for decay
of the off diagonal elements in Eqs. (\ref{eqn:solnHB}) and
(\ref{eqn:solnLB}), with equality only when $\theta=0$. This
indicates that the detector is inefficient unless the qubit energy
eigenstates are localized.

\section{Quantum Trajectories}

\begin{figure*}
\subfigure[]{\includegraphics[width=4cm]{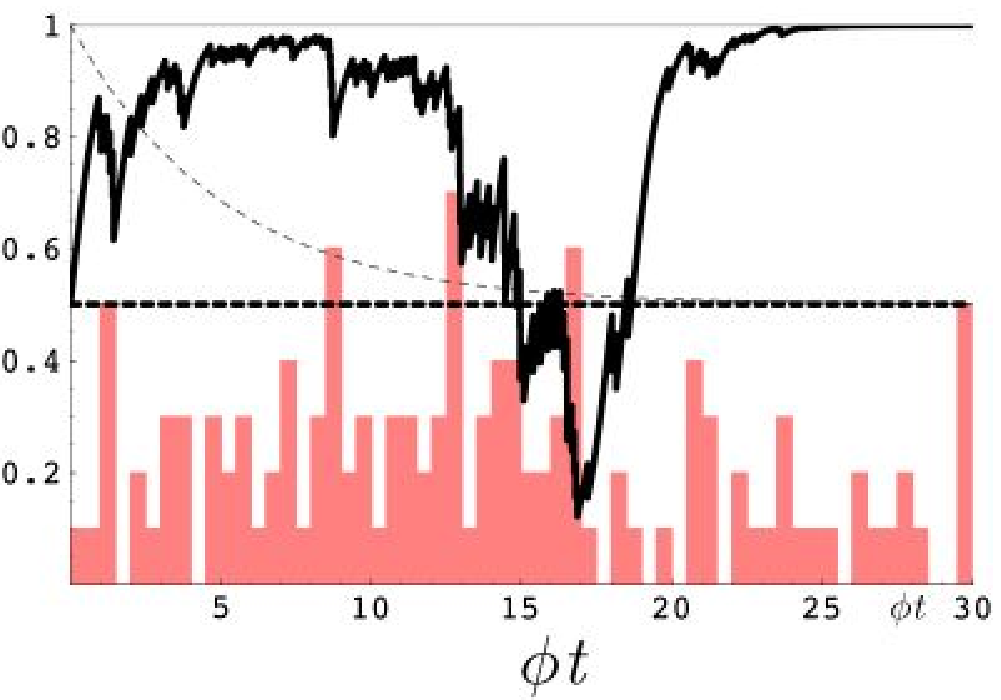}\label{fig:trajectory0l}}
\subfigure[]{\includegraphics[width=4cm]{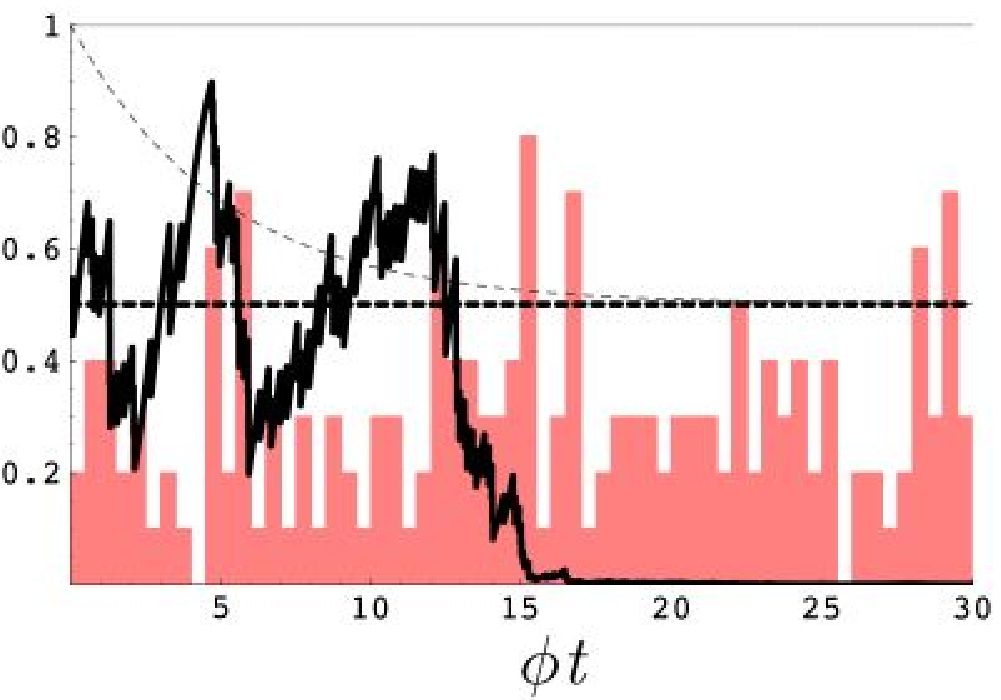}\label{fig:trajectory0r}}
\subfigure[]{\includegraphics[width=4cm]{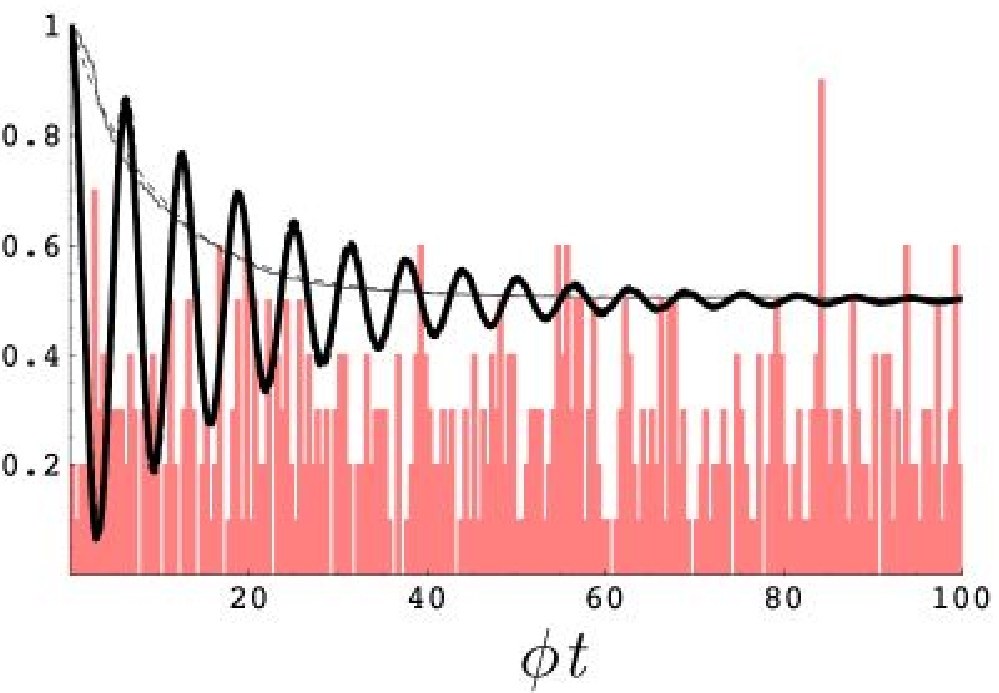}\label{fig:TrajectoryV05th1}}
\subfigure[]{\includegraphics[width=4cm]{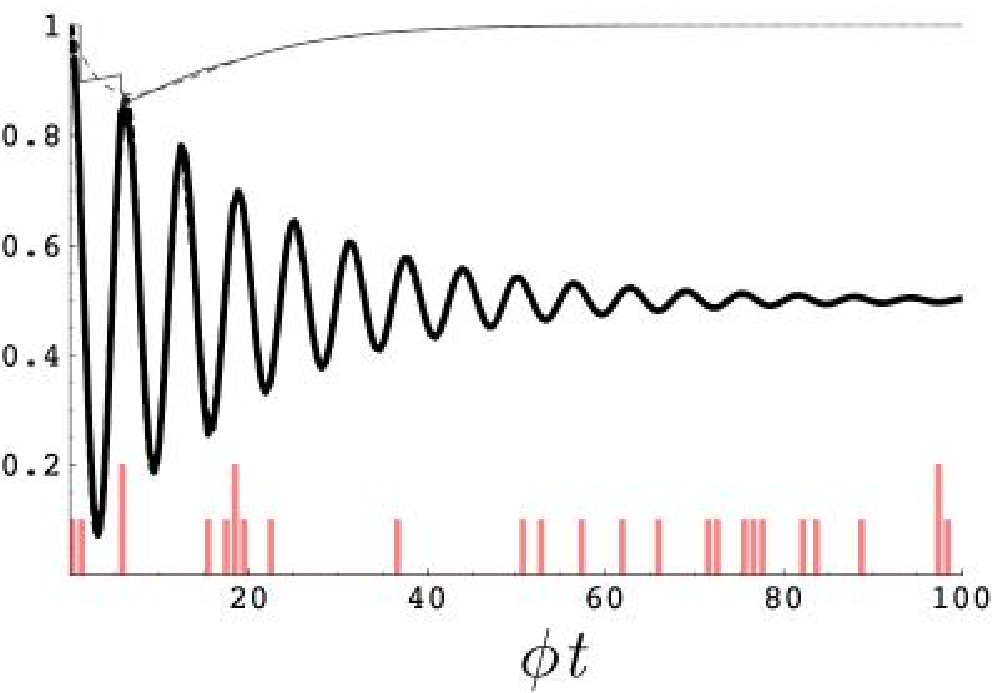}\label{fig:trajectory2}}
\caption{\label{fig:degeneratetrajectories} (a) \& (b)  Showing
distinct simulations where the qubit collapses to $\ket{l}$ and
$\ket{r}$ respectively [for $\theta=0, e V/\splitting=10,
\nu^2=0.005, \Tnd^2=0.5$].  At $\theta=\pi/2$ currents are
uncorrelated with the qubit in either (c) high-bias [$e
V/\splitting=10, \nu^2=0.005, \Tnd^2=0.5$]  or (d) low-bias   [$e
V/\splitting=0.5, \nu^2=0.05, \Tnd^2=0.5$].  In all panels, dark
curves are $\bra{l}\rho_c(t)\ket{l}$  (solid) and
$\bra{l}\rho(t)\ket{l}$ (dashed), and light curves are the
conditional purity $\tr{\rho_c(t)^2}$ (solid) and unconditional
purity $\tr{\rho(t)^2}$ (dashed).  Histograms show the  number of
jumps per time interval, (scaled by $1/10$).}
\end{figure*}

In \figs{fig:trajectory0l} and \ref{fig:trajectory0r} we show two
different quantum trajectory simulations for $\theta=0$, using the
same parameters to generate both (see caption for values).  We
show the conditional probability (dark, solid line),
$\bra{l}\rho_c(t)\ket{l}$, and unconditional probability (dark,
dashed line), $\bra{l}\rho(t)\ket{l}$, of the qubit to be in the
left well.  Also shown is the purity of conditional state (light,
solid), $\tr{\rho_c(t)^2}$, and of the unconditional state
$\tr{\rho(t)^2}$, (light, dashed).  We also show a histogram of
the number of jump processes that were encountered in each time
window, and this corresponds to experimentally measured currents.
In all trajectories shown in this paper, the histograms are scaled
by the factor $\frac{1}{10}$, and are mutually comparable.  In
these two figures, the initial state of the system is chosen to be
the pure state $(\ket{l}+\ket{r})/\sqrt{2}$.

The evolution of the measurement clearly shows the qubit
collapsing to the state $\ket{l}$ or $\ket{r}$, and the state
remains pure throughout.  This is because, at $\theta=\pi/2$, the
inelastic jump operators $c_1$ and $c_2$ are suppressed, and the
remaining jump operator $c_3$ partially projects the state onto
the localized basis.  The average current, determined by the mean
of the histogram is higher in the case where the qubit collapsed
onto the state $\ket{r}$, which is consistent with physical
intuition, since in this configuration, the tunnelling rate
through the PC is highest.  The fact that the unconditional
probability to find the electron on the left well is constant at
0.5 reflects the fact that the state of the system collapses to
$\ket{l}$ or $\ket{r}$ with equal probability, consistent with the
initial state preparation being an equal superposition of the
localized states, and thus for $\theta=0$, the PC serves as a good
quantum non-destructive (QND) measurement device.

Figures \ref{fig:trajectory2} and \ref{fig:TrajectoryV05th1} shows
the dynamics of the system for $\theta=\pi/2$ in the high bias ($e
V/\splitting=2$) and low bias ($e V/\splitting=0.5$) regimes,
respectively.  In both cases, the initial state of the system is
$\ket{l}$.

As predicted by \eqn{eqn:solnHB}, in the high bias limit the
equilibrium state of the system is an unequal mixture of $\ket{g}$
and $\ket{e}$.  We may calculate the unconditional steady state
purity, and we find
$\tr{\rho(\infty)}=\frac{(eV)^2+\splitting^2}{(2 e V)^2}=0.505$
for the parameters used.  This agrees with the simulation of
\fig{fig:trajectory2}, where the conditional and unconditional
evolution is very similar.

Similar comments apply to \fig{fig:TrajectoryV05th1}, however the
unconditional steady state is always $\ket{g}$ for the low bias
limit, which is of course a pure state, and for the case
considered is just an equal superposition of localized states
$\ket{g}=(\ket{l}+\ket{r})/\sqrt{2}$.  Thus the probability for
the qubit to be in the left well is just the same as for
\fig{fig:trajectory2}, but the equilibrium state is pure.

\begin{figure*}
$\theta=\pi/8$\hspace{4cm}$\theta=\pi/4$\hspace{4cm}$\theta=3\pi/8$\\
{\includegraphics[angle=90]{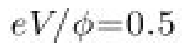}}
{\includegraphics[width=5cm]{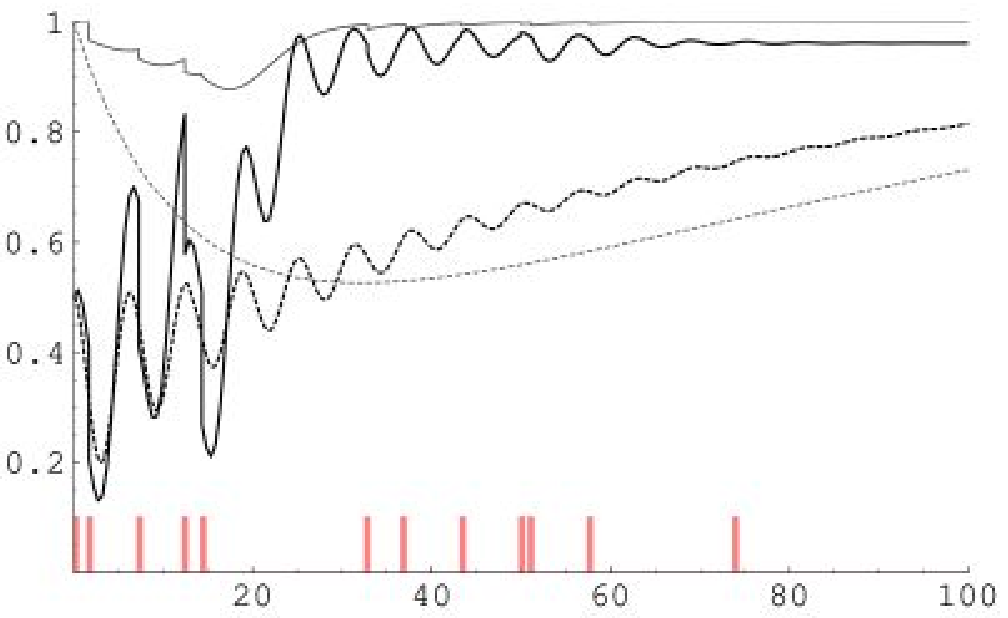}\label{fig:TrajectoryV05th4}}
{\includegraphics[width=5cm]{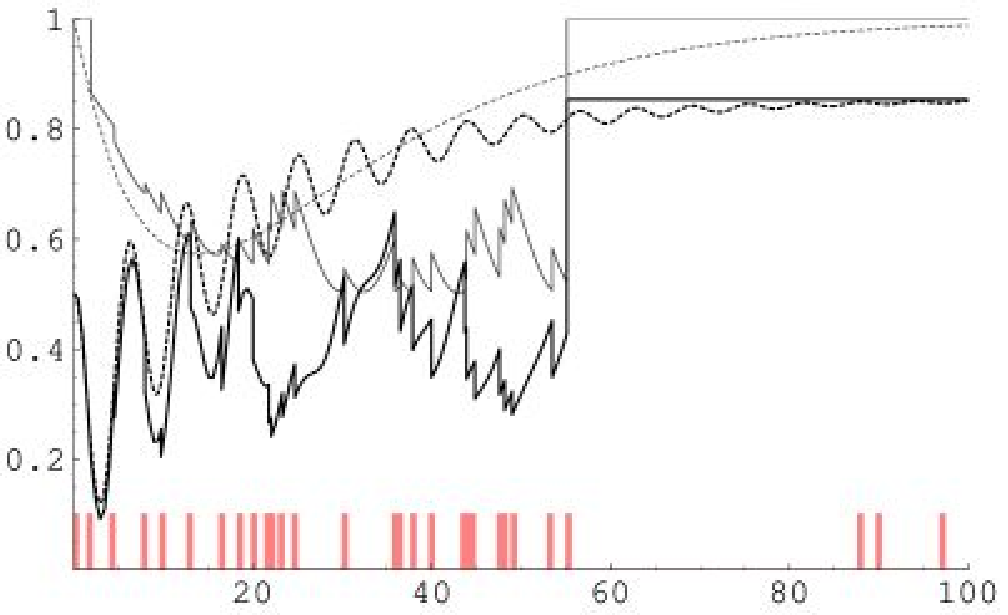}\label{fig:TrajectoryV05th2}}
{\includegraphics[width=5cm]{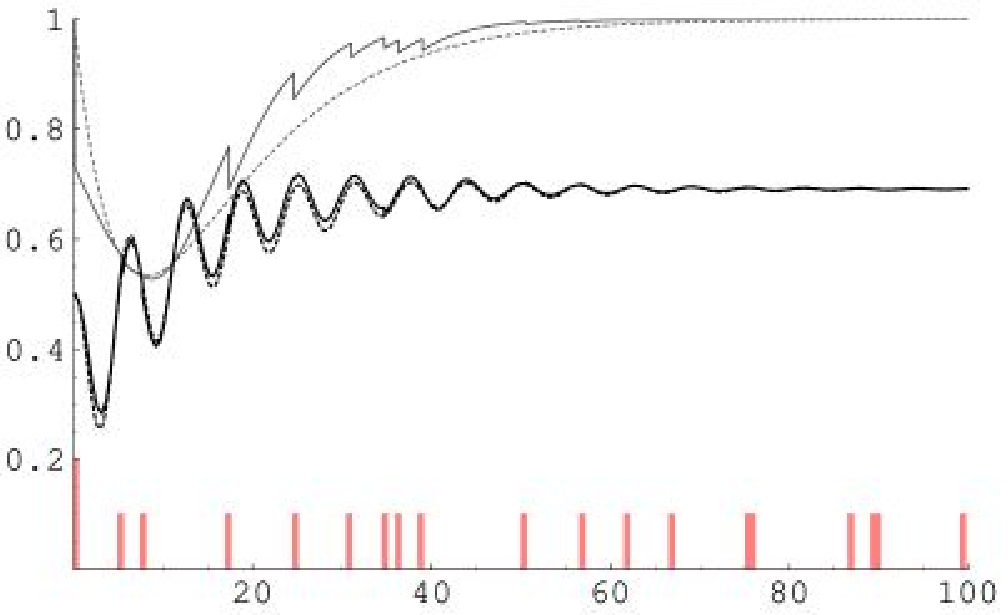}\label{fig:TrajectoryV05th34}}\\
{\includegraphics[angle=90]{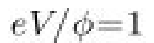}}
{\includegraphics[width=5cm]{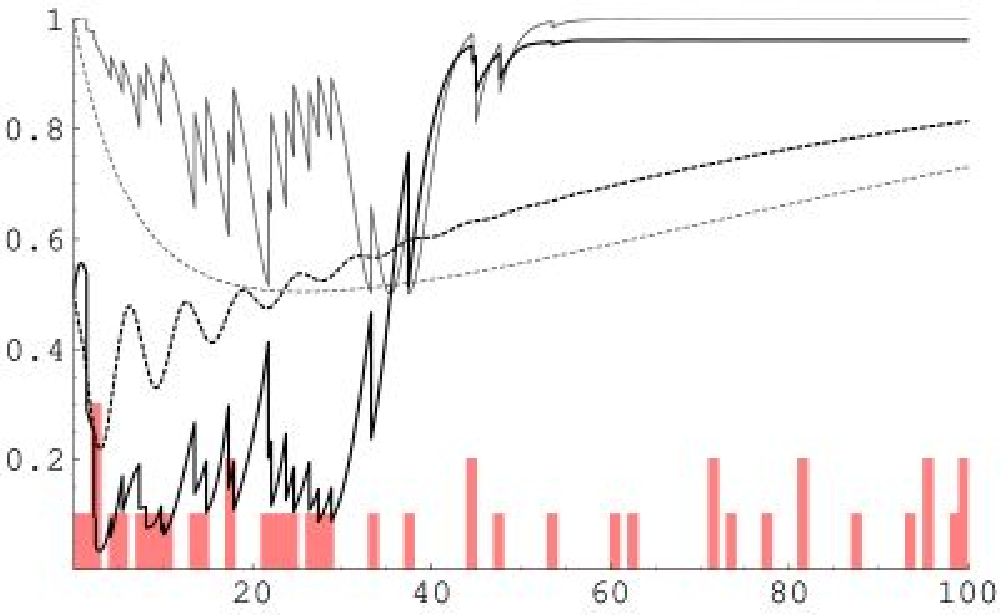}\label{fig:TrajectoryV09th4}}
{\includegraphics[width=5cm]{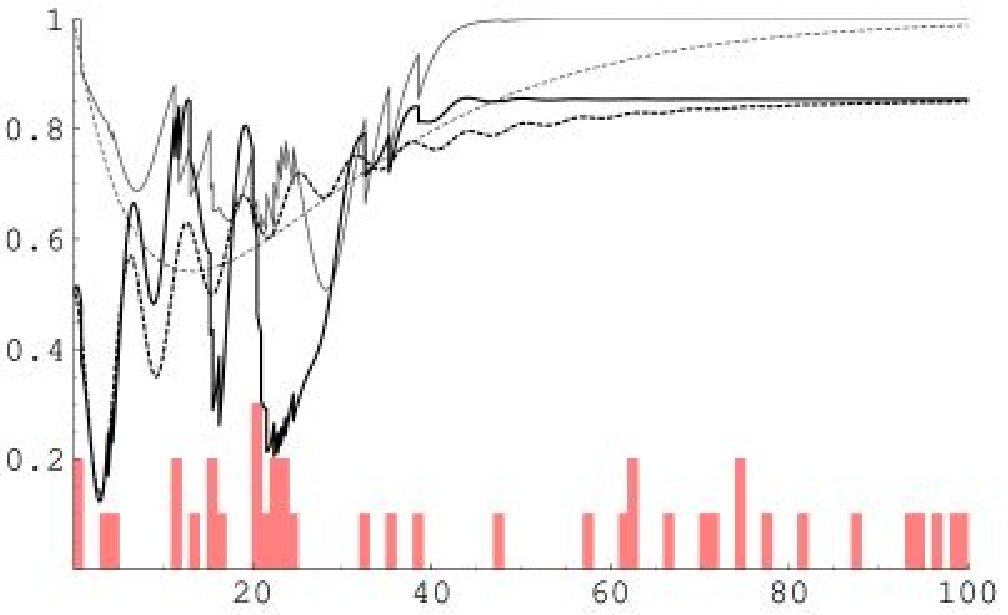}\label{fig:TrajectoryV09th2}}
{\includegraphics[width=5cm]{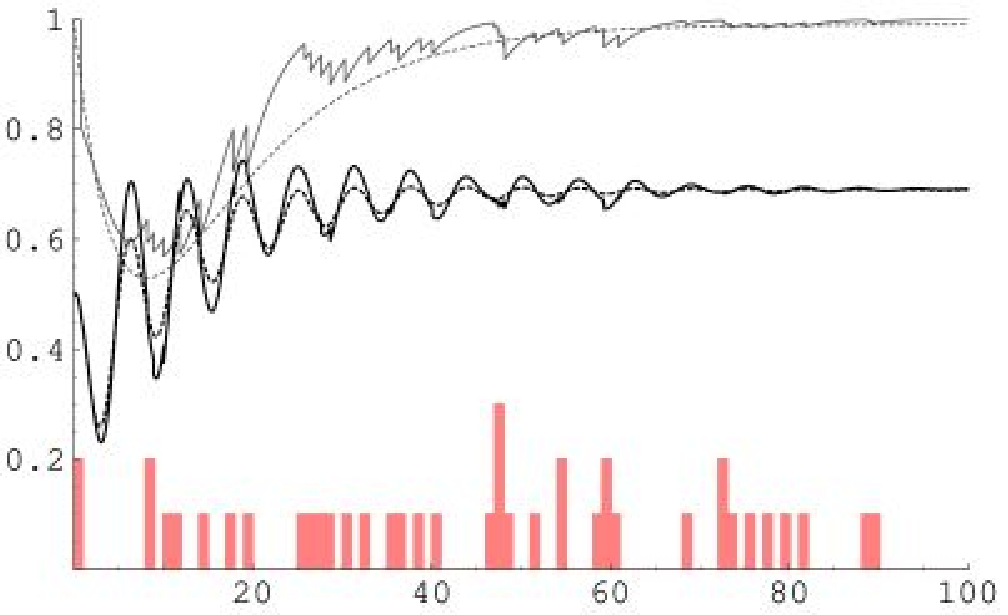}\label{fig:TrajectoryV09th34}}\\
{\includegraphics[angle=90]{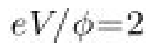}}
{\includegraphics[width=5cm]{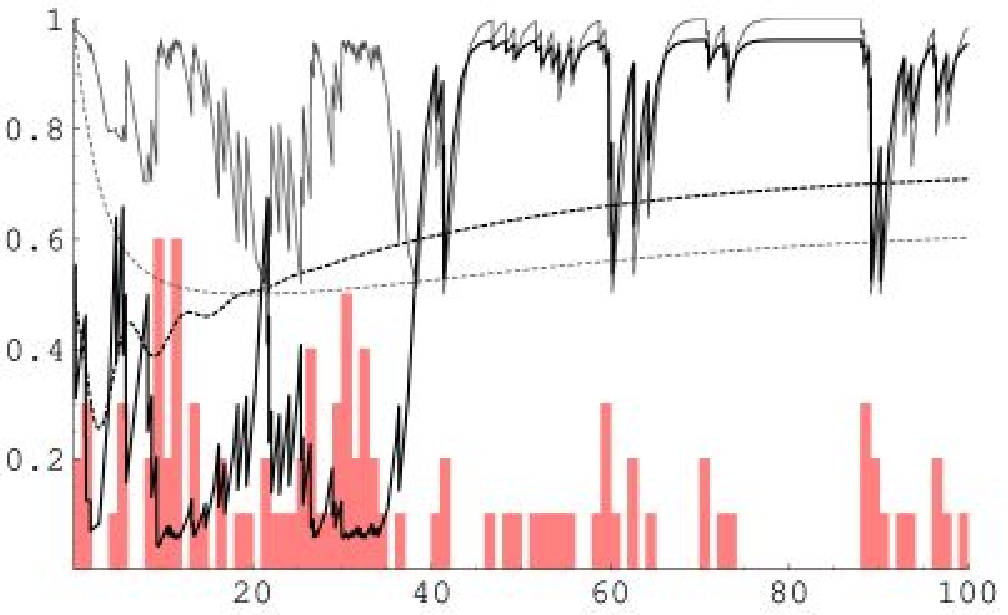}\label{fig:TrajectoryV11th4}}
{\includegraphics[width=5cm]{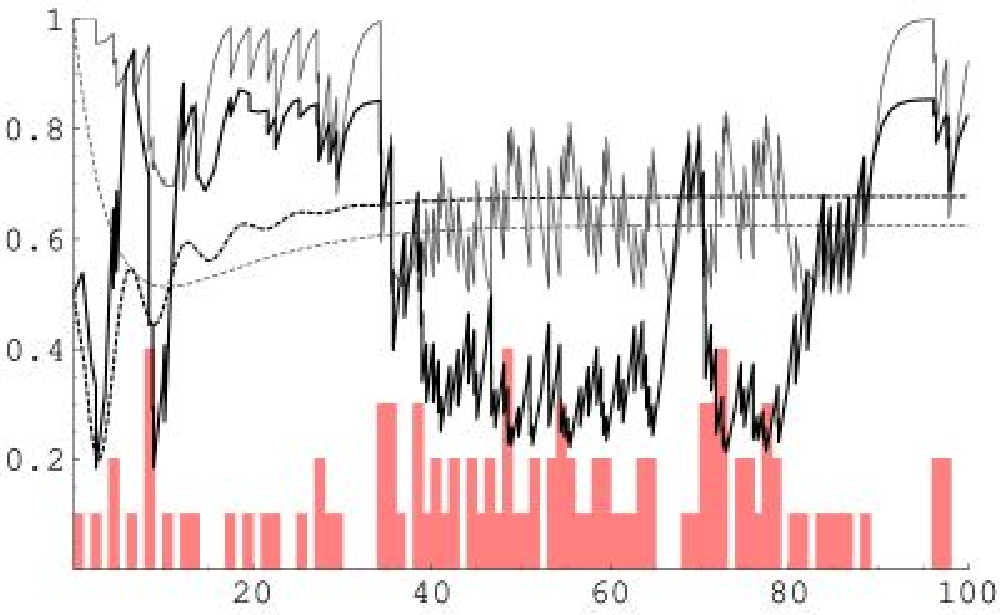}\label{fig:TrajectoryV11th2}}
{\includegraphics[width=5cm]{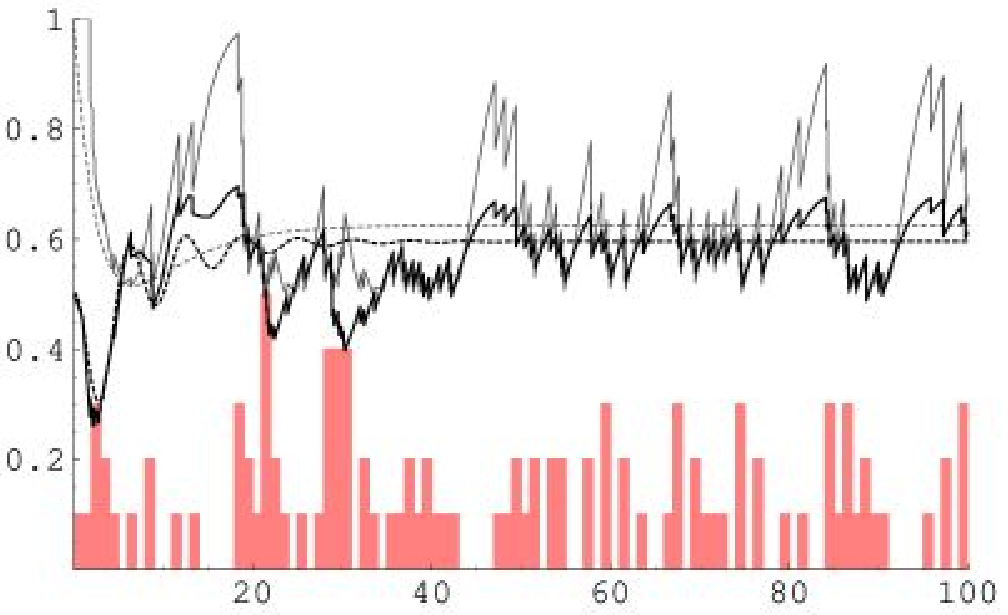}\label{fig:TrajectoryV11th34}}\\
{\includegraphics[angle=90]{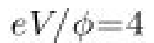}}
{\includegraphics[width=5cm]{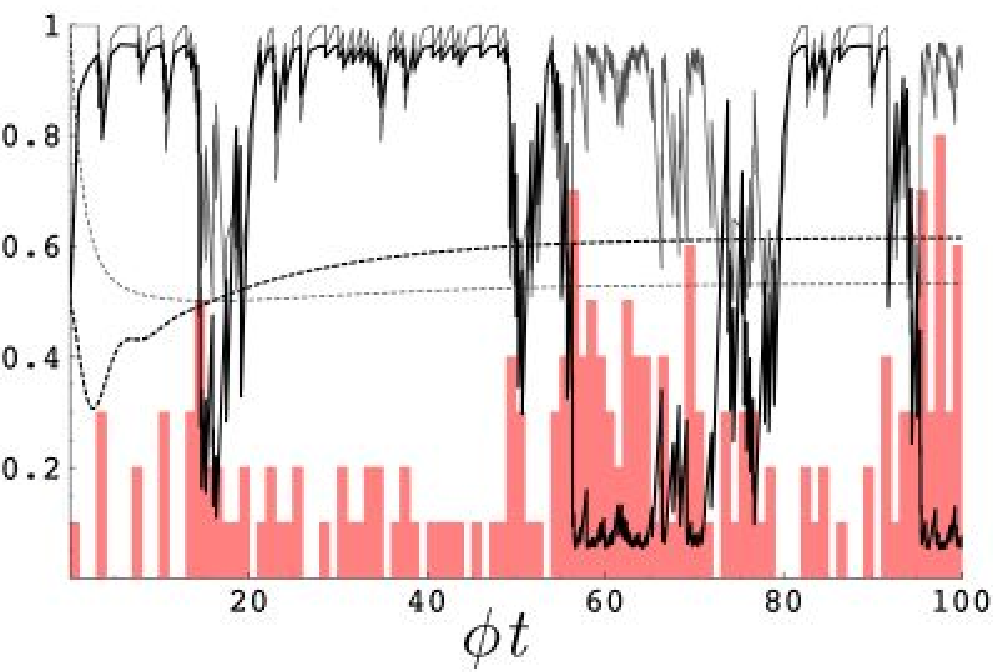}\label{fig:TrajectoryV2th4}}
{\includegraphics[width=5cm]{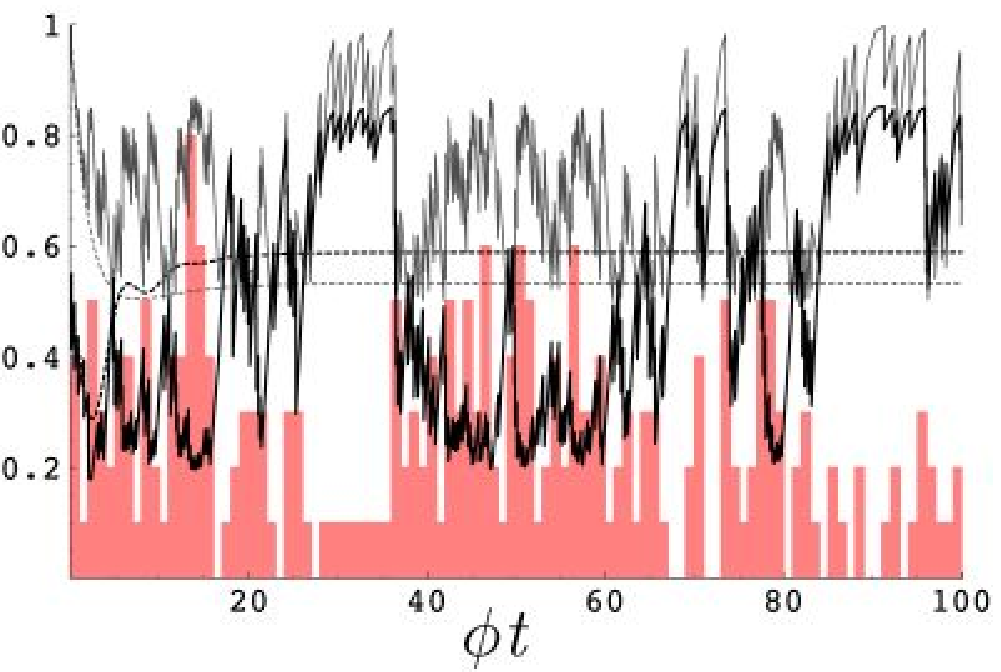}\label{fig:TrajectoryV2th2}}
{\includegraphics[width=5cm]{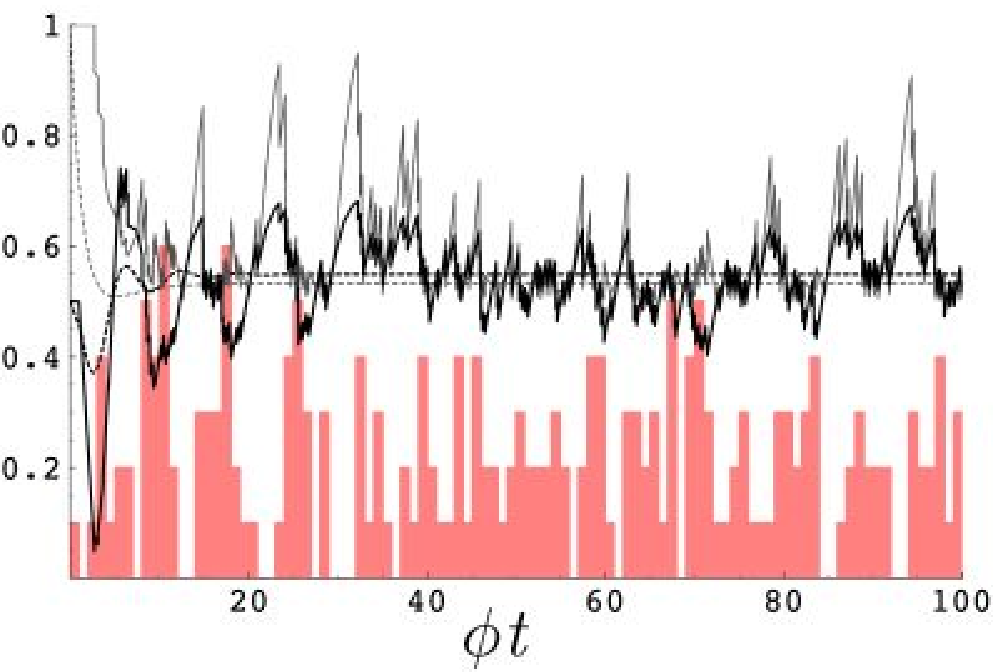}\label{fig:TrajectoryV2th34}}
\caption{\label{fig:ParamStudy}Simulated trajectories from left to
right, $\theta=\pi/8, \pi/4$ and $3\pi/8$.  From top to bottom, $e
V/\splitting=0.5, 0.9, 1.1,  2$ and 4.  In all $\nu^2=0.05$ and
$\Tnd^2=0.5$.  In each panel, the dark curves are
$\bra{l}\rho_c(t)\ket{l}$ (solid) and $\bra{l}\rho(t)\ket{l}$
(dashed), and the light curves are the purity of the conditional
state $\tr{\rho_c(t)^2}$ (solid) and the unconditional state
$\tr{\rho(t)^2}$ (dashed).  The histogram shows the number of
jumps that occurred in each time interval.  The histograms are all
scaled by the same factor, so are comparable with one another.
Reverse tunnelling events are responsible for the sharp changes in
the first two panels.}
\end{figure*}

We illustrate the effect of changing parameters in
\fig{fig:ParamStudy}.  We show simulated trajectories for five
different bias voltages, $e V=0.5, 1, 2$ and 4, and for three
different mixing angles $\theta=\pi/8, \pi/4$ and $3\pi/8$.  In
the high bias regime, and small values of $\theta$ there are
fluctuations that bear a superficial resemblance to the quantum
Zeno effect (QZE).  The state of the qubit tends to fluctuate
between mixtures of energy eigenstates, and the conditional
density matrix is diagonal in the energy eigenbasis, with
off-diagonal terms being negligible in this basis.

This resemblance to the QZE is only superficial however.  In the
QZE, the free dynamics of the qubit are suppressed due to the
strong and frequent measurement which project the system into
pure, localized states $\{\ket{l},\ket{r}\}$.  The sharp
transitions that occur in the QZE  result from the small
probability for the electron to make a transition between sites
which is realized as a rapid, occasional tunnelling event.

In contrast, the transitions that are evident in
\fig{fig:ParamStudy} for small $\theta$ are due to relaxation or
excitation of the qubit, accompanied by inelastic tunnelling
processes through the PC, rather than the `collapse of the
wavefunction' due to accumulated information about the qubit in
the case of the QZE.  As $\theta$ increases, the transitions
become smaller in amplitude and less distinct, until
$\theta=\pi/2$ when  inelastic tunnelling events through the PC
dominate the current and the qubit state is damped to the
completely mixed state.

At the transition from low bias to high bias, $e V/\splitting=1$,
fluctuations begin to appear in the steady state trajectory.  For
$e V<\splitting$ the conditional steady state of the qubit is
constant, whereas for $e V>\splitting$ the conditional steady
state shows fluctuations which increase with increasing bias
voltage.

\section{Discussion}

The predictions of this work should be experimentally verifiable.
In particular, it should be possible, in near-future experiments,
to observe the qualitative change in both the steady state
current, and the fluctuations in the current, in the transition
from the low- to high-bias regimes. This sharp transition between
low- and high-bias regimes also provides possible techniques for
performing spectroscopy of solid state qubits, as discussed in
Sec. \ref{sec:Iss}.

The results presented here describe the dynamics of the measured
qubit in the sub-Zeno limit, $\Gamma_d\ll\splitting$, for
arbitrary PC bias $eV$, whereas previous analyses
\cite{goa01a,kor01a,kor01b,gur03} are valid only for large PC
bias, $eV\gg\splitting$, but arbitrary $\Gamma_d / \splitting$.
Outside these regimes, for finite ratios $\Gamma_d / \splitting$
and $eV / \splitting$  neither approach is formally valid, and
non-Markovian effects may play a significant role.  This parameter
regime is therefore an open area for investigation.

There are a number of practical issues that the present paper
raises.  Firstly, as discussed earlier, to turn the measurement
off, it is not enough just to turn the PC bias to zero.
Simultaneously one must also make $\nu^2$ small, which can be
accomplished with extra surface gates.

Secondly, in order to perform a good single-shot measurement we
require that the qubit measurement time is much shorter than the
detector-induced relaxation time,\cite{mak01} i.e.
$\tmeas\ll\trel$. From \eqns{eqn:solnHB}, (\ref{eqn:solnLB}) and
(\ref{eqn:dz2}) we see that $\trel^{-1}=2\nu^2 \max(\splitting,e
V) \sin^2(\theta)$ and $\tmeas^{-1}=2 \Gamma_d\cos^2(\theta)$, so
we require that $\sin(\theta)\approx0$.  Therefore, our work
indicates that measurements in the sub-Zeno regime are not
possible for highly delocalized qubit eigenstates.  Furthermore,
ideal measurements (i.e. measurements for which the conditional
state remains pure) are only attainable when $\theta=0$, so the
energy eigenstates are the localized states.

In order to perform good single-shot measurements, one could
operate in the Zeno regime, though there may be technical
difficulties in obtaining a sufficiently large value of
$\Gamma_d$, since this requires passing large currents through the
PC. These large currents can, in turn, lead to heating in the
detector, which can significantly increase noise and lead to
additional decoherence of the qubit.\cite{gar03,SeansThesis}

Alternatively, performing a good single-shot measurement in the
sub-Zeno regime consists of three tasks:
\begin{enumerate}
\item Localise the qubit eigenstates by turning on the qubit bias
($\epsilon$) and/or decreasing qubit tunnelling rate ($\Delta$),
so $\theta\approx0$. \item Increase the lead tunnelling rate
($\too$ and $\xoo$) by lowering a tunnel barrier. \item Turn on
the lead bias voltage ($V$).
\end{enumerate}
The rate at which $\theta$ is varied effects a rotation on the
qubit, so by selecting this rate appropriately, we can choose to
measure  in an arbitrary basis.

\section{Conclusion}

In this paper we have used the quantum trajectories formalism to
derive master equations for measurements of a charge qubit by an
external point contact electrometer in the sub-Zeno regime.  The
master equation was derived without recourse to heuristic
arguments, and resulted in the inclusion of inelastic processes in
which energy is exchanged between the qubit and the detector.
These inelastic processes have a profound effect on the
conditional and unconditional dynamics of the system. Furthermore,
within the sub-Zeno regime, our results are valid for arbitrary
detector bias voltage.

In the low-bias regime ($e V <\splitting$) the qubit always
relaxes to its ground state, much like a qubit in equilibrium with
a zero-temperature bath.  In this case, relaxation to the ground
state is due to the eventual spontaneous relaxation of the qubit
accompanied by an excitation of a PC lead electron.  The
corresponding steady state power spectrum is flat.

In the high-bias regime ($e V>\splitting$) the PC leads act like a
zero-entropy heat bath at non-zero temperature and both inelastic
relaxation and excitation processes take place, causing the steady
state of the qubit to be a mixture of excited and ground states.
These inelastic transitions are also reflected in the power
spectrum, which exhibits a peak centered at zero frequency,
corresponding to transitions of the qubit between the localized
energy eigenstates.

The sharp transition the dynamics at $e V = \splitting$ is also
reflected in the steady state conductance of the PC. This
observation provides techniques for accurately determining the
qubit Hamiltonian parameters directly from conductance
measurements.

Single shot measurement remains possible in the sub-Zeno regime,
provided that the system Hamiltonian is modified such that the
energy eigenstates become localized before the measurement takes
place. This also gives the added freedom that we may choose an
arbitrary basis in which to measure.

TMS thanks the \tms{Hackett committee, the CVCP and Fujitsu} for
financial support.  SDB acknowledges support from the E.U.
NANOMAGIQC project (Contract no. IST-2001-33186).  We thank H.-S.\
Goan, W.\ J.\ Munro, T.\ Spiller, G.\ J.\ Milburn, H-A.\ Engel,
R.\ Aguado, A.\ Shnirman, H.\ M.\ Wiseman and D.\ Averin for
useful conversations.  In particular, H.\ M.\ Wiseman suggested
\fig{fig:RegionOfValidity}.

\appendix

\section{Generalized power spectrum of Goan and Milburn \cite{goa01a}}\label{app:GMPS}

The power spectra calculated in \cite{goa01a} assumed
$\epsilon=0$.  In order to compare our power spectra with theirs
for arbitrary parameter values, we derive the power spectra for
their model for arbitrary $\epsilon$.

 It is laborious to compute $\ggm(\tau)$ directly, however since we are only interested in the power spectrum, we can bypass the explicit solution for the correlation function in the time domain and calculate the power spectrum directly.

 We wish to compute the Fourier transform, $\FT$, of equation (A8) of  \citep{goa01a} (hereafter called (GMA8)).  This may be done via a Laplace transform, $\LT$, using the relation for a symmetric function $f(t)$
 \begin{equation}
 \FT_\omega[f(t)]=\LT_{i \omega}[f(t)]+\LT_{-i \omega}[f(t)].
 \end{equation}
 The Laplace transform of (GMA8) is
 \begin{equation}
\LT_s[\ggm(\tau)]=e \,\iloc+\diloc^2(\Tr\{n_1 \LT_s[e^{\Lind
\tau}n_1 \rho_\infty]\}-\Tr\{n_1
\rho_\infty\}^2/s).\label{eqn:LTG1}
\end{equation}
The notation $e^{\Lind \tau}n_1 \rho_\infty$ is just shorthand for
the solution to the unconditional master equation (GM5a) at time
$\tau$ subject to the initial condition $\rho(0)=n_1
\rho_\infty=n_1/2$.  Taking the Laplace transform of (GM5a) gives
\begin{equation}
s\LT_s[\rho(t)]-\rho(0)=-i\commute{\hsys}{\LT_s[\rho(t)]}+\lind[T_l+(T_r-T_l)
n_1]\LT_s[\rho(t)].
\end{equation}
$\LT_s[\rho(t)]$ may be found straightforwardly from this
expression and then $\LT_s[e^{\Lind \tau}n_1 \rho_\infty]$ is
obtained by substituting $\rho(0)\rightarrow n_1
\rho_\infty=n_1/2$ into  $\LT_s[\rho(\tau)]$, though it is a
somewhat cumbersome expression.  Evaluating the traces in
\eqn{eqn:LTG1} yields
 \begin{equation}
\LT_s[\ggm(\tau)]=\frac{e\, \iloc}{2}+\diloc^2\frac{4s^2 +
2{\splitting }^2 + 4s{\xgm }^2 + {\xgm }^4 + 2{\splitting }^2\cos
(2\theta )}
  {4\left( 4s^3 + 4s^2{\xgm }^2 + {\splitting }^2{\xgm }^2 +
      s\left( 4{\splitting }^2 + {\xgm }^4 \right)  - {\splitting }^2{\xgm }^2\cos (2\theta )
      \right) },\label{eqn:LTG}
\end{equation}
where we have used the same definition of $\xgm$ as found in
\citet{goa01a}, which for comparison is given in our notation as
$\xgm=\nu \sqrt{e V}$.
 The power spectrum is then given by
 \begin{equation}
 \sgm(\omega)=2\,\FT_\omega[\ggm(t)]=2(\LT_{i \omega}[\ggm(\tau)]+\LT_{-i\omega}[\ggm(\tau)])
 \end{equation}

\renewcommand{\url}[1]{}
\newcommand{\urlprefix}[1]{#1}

\end{document}